\documentclass[preprint2]{aastex}

\newcommand{\etal}{{\it et al.~}}

\newcommand{\bpz}{{\it BPZ}}
\newcommand{\sed}{{\it SED}}

\newcommand{\sn}{{S/N}}
\newcommand{\dls}{{DLS}}
\newcommand{\zp}{{z_{\rm phot}}}
\newcommand{\zs}{{z_{\rm spec}}}
\newcommand{\zi}{{z_{\rm input}}}
\newcommand{\dz}{{\Delta z}}

\begin{document}

\title{Photometric Redshifts and Signal-to-Noise}

\author{V. E. Margoniner\altaffilmark{1} and D. M. Wittman\altaffilmark{1}}

\altaffiltext{1}{Department of Physics, University of California at
Davis, One Shields Avenue, Davis, CA 95616}

\begin{abstract}

We investigate the impact of photometric signal-to-noise (\sn) on the
precision of photometric redshifts in multi-band imaging surveys,
using both simulations and real data.  We simulate the optical 4-band
(BVRz) Deep Lens Survey (\dls, Wittman \etal 2002), and use the
publicly available Bayesian Photometric Redshift code \bpz~ by Benitez
(2000).  The simulations include a realistic range of magnitudes and
colors and vary from infinite S/N to $S/N=5$.  The real data are from
DLS photometry and two spectroscopic surveys, and explore a range of
S/N by adding noise to initially very high S/N photometry.  Precision
degrades steadily as S/N drops, both because of direct S/N effects and
because lower S/N is linked to fainter galaxies with a weaker
magnitude prior.  If a simple S/N cut were used, $S/N\ge 17$ in
R (corresponding, in the DLS, to lower S/N in other bands) would be
required to keep the scatter in $\dz\equiv {\zs - \zp \over 1 + \zs}$
to less than 0.1.  However, cutting on ODDS (a measure of the
peakiness of the probability density function provided by BPZ) greater
than 0.4 provides roughly double the number of usable galaxies with
the same $\sigma_{\dz}$.  Ellipticals form the tightest $\zs-\zp$
relation, and cutting on type=elliptical provides better precision
than the $ODDS>0.9$ cut, but this eliminates the vast majority of
galaxies in a deep survey.  In addition to being more efficient than a
type cut, ODDS also has the advantages of working with all types of
galaxies (although ellipticals are overrepresented) and of being a
continuous parameter for which the severity of the cut can be
adjusted as desired.

\end{abstract}


\section{Introduction}\label{introduction}

Photometric redshifts (Connolly \etal~1995, Hogg \etal~1998, Benitez
2000) are of paramount importance for current and planned multi-band
imaging surveys.  With photometric redshifts, surveys can
inexpensively gather information about structure along the line of
sight, without resorting to expensive spectroscopic followup.
Therefore, it is important to understand systematic errors and
limitations in this method.  For example, Ma \etal~2006 and Huterer
\etal~2006 have examined the required photometric redshift accuracy
for surveys which plan to use weak lensing (cosmic shear) to constrain
dark energy.  For this application and also for baryon acoustic
oscillations (Zhan \& Knox 2006), reducing photometric redshift errors
is less important than knowing the error distribution accurately.
Thus, careful attention must be paid to systematic differences between
the photometric survey and the spectroscopic sample used to evaluate
photometric redshift performance.  For most surveys, photometric S/N
is one of the systematic differences.

The most well-known test case for photometric redshifts is the blind
test in the Hubble Deep Field North (HDFN) conducted by Hogg \etal\
(1998).  The best methods then yielded $\sigma_{\dz}\sim0.1$, where
$\dz\equiv {\zs - \zp \over 1 + \zs}$, using Hubble Space Telescope
(HST) photometry in UBVI bands and ground JHK (Dickinson ~1998). More
recently, with improved photometry and spectral redshift
classification, an accuracy of $\sigma\dz\sim0.06$ is achieved over
the redshift range 0---6 (Fernandez-Soto \etal~1999, 2001; Benitez
2000).  Ground-based surveys suffer from less precise photometry but
usually do not have to deal with such a large redshift range.  Ilbert
\etal~2007 cite an accuracy of $\sigma_{\dz} = 0.029$ after clipping
outliers with $\dz>0.15$ (3.8\% of the sample).  Ilbert \etal~2007
also find a decrease in precision at fainter magnitudes, but made no
effort to separate the effects of S/N from the other effects operating
on faint galaxies, such as a weaker magnitude prior and greater SED
evolution.  In this paper, we examine the impact of these effects
separately, focusing on photometric S/N.  
%We use both simulations and
%a real spectroscopic dataset much larger than that of the HDF.
The quantitative results presented here are specific to the BVRz
filter set used in the Deep Lens Survey (\dls, Wittman \etal~2002).
More filters, covering a wider range in wavelengths, will do better
(Abdalla \etal~2007).  However, the trends with S/N are broadly
applicable.

\section{Method}\label{method}

We use the \bpz~ Bayesian photometric redshift code developed by
Benitez (2000).  We also tested the HyperZ code (Bolzonella \etal\
2000) with additional priors roughly equivalent to the default BPZ
priors, and found similar performance.  For clarity we present only
results from BPZ here.  We did not test training-set methods, in which
a spectroscopic and photometric training set is used to perform a fit
or to train a neural network, for two reasons.  First, training set
methods are unlikely to be employed for surveys planning to push the
photometric sample deeper than the spectroscopic sample.  Second, the
two methods seem to be roughly equivalent in performance on the data
sets in which they have been compared (e.g. Hogg \etal~1998), so the
trends presented here should be applicable to both methods.

We use the six spectral energy distribution (\sed) templates from
Benitez (2000): E, Sbc, Scd, Irr, SB3, and SB2, modified as described
below.  For the simulations, the same templates are used to simulate
the photometry and to infer the photometric redshifts; there is no
allowance for cosmic variance of the templates or ``template noise''.
For the data, it is important that the templates reflect real SEDs.
Therefore, we use the photometry of objects with spectroscopic
redshifts to optimize the templates (Csabai \etal~2000; Benitez
\etal~2004; Ilbert \etal~2007). Section~\ref{tempopt} describes the
procedure and shows the corrected templates.  Clearly, even the
optimized templates do not represent all types of SEDs in the
universe. For both simulations and data, we start by demonstrating the
performance with as nearly perfect a data set as possible.  After
illustrating the best-case scenarios, we proceed to degrade the
simulations and data to successively lower S/N, repeating the analysis
for each step.

For each galaxy, we identify the peak of its redshift probability
density function (PDF) as its {\it photometric redshift} or $z_{\rm
phot}$.  This greatly simplifies the analysis and presentation of the
results, at the cost of some precision.  Specifically, ``catastrophic
outliers'' will appear, whose $z_{\rm phot}$ differs greatly from
their true redshift.  In many cases, this may be an artifact of not
considering the full PDF, a point argued forcefully in the case of the
HDF by Fernandez-Soto
\etal (2001, 2002).  The full PDF may contain additional peaks, or
otherwise be broad enough to be consistent with the true redshift.  In
this paper, we wish to focus on the trends with photometric S/N rather
than the characterization of outliers.  As will be seen in the tables
and figures, the trends with S/N are not substantively changed if
``outliers'' are removed.  Therefore we judge this simplification to
be acceptable.  ``Outlier'' in this paper thus refers to difference
between $z_{\rm phot}$ and true redshift, without implying anything
about the full PDF.

We do consider characteristics of the PDF when using BPZ's ODDS
parameter.  BPZ assumes a natural error (template noise) of
$0.067(1+z)$, and defines ODDS as the fraction of the area enclosed by
the PDF between $zphot\pm n\times0.067(1+z)$, where $n$ is a
user-settable parameter which we set to 1.  ODDS values close to unity
indicate that most of the area under the redshift probability density
function (PDF) is within $Z_B\pm0.067(1+z)$.  In this paper, we
present results both for the entirety of a given sample, and after a
cut of $ODDS>0.9$, which eliminates many of the ``outliers.''  We
also investigate the tradeoff between ODDS cut, number of usable
galaxies, and photometric redshift accuracy.

The error distributions are typically non-Gaussian, often highly so.
The rms or standard deviation is extremely sensitive to even a few
non-Gaussian events, so in the photometric redshift literature,
results are usually quoted as an rms after excluding a certain (small)
fraction of galaxies as ``catastrophic outliers.''  The fraction
varies from paper to paper, making comparison difficult.  The field of
robust statistics suggests several less sensitive metrics of
variation, such as the median or mean absolute deviation.  However,
outliers {\it should} be included in the performance analysis with
some weight, because they will be included when using the entire
photometric sample for science.  We therefore clip conservatively,
$|\dz|<0.5$, to avoid overly optimistic results.  This threshold
is at least five, and usually many more, times the clipped rms.  We
also present, in many cases, differential and cumulative distributions
as well.  To make the connection with forecasts for, say, weak lensing
tomography, we suggest these distributions be fit with double
Gaussians.  Gaussians are analytically tractable, and a double
Gaussian can fit both the core and wings (but not truly catastrophic
outliers).

%-----------------------------------------------------------------%
%-----------------------------------------------------------------%

\section{Simulations}\label{simulations}

\begin{figure}
%\epsscale{.80}
\plottwo{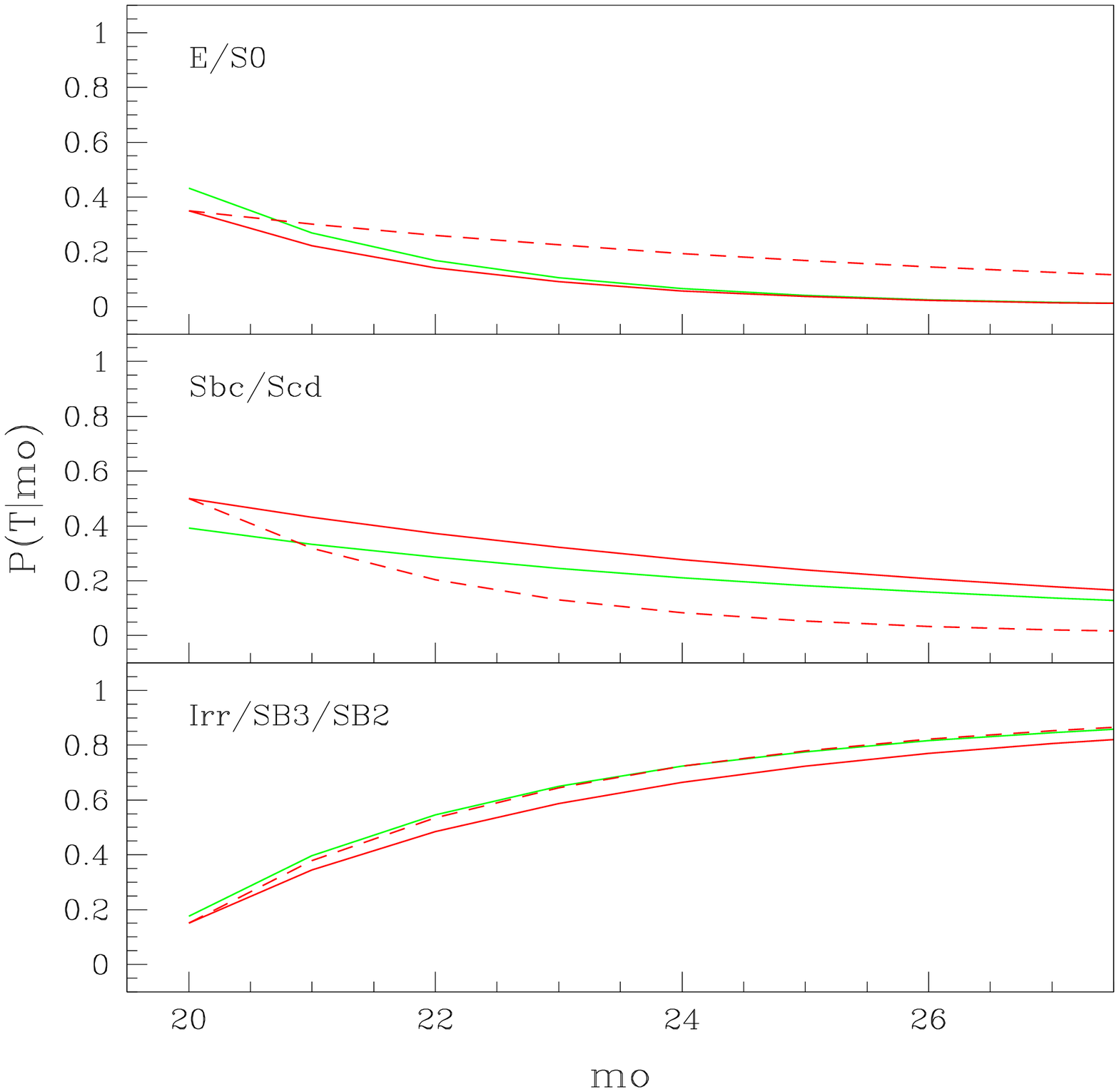}
        {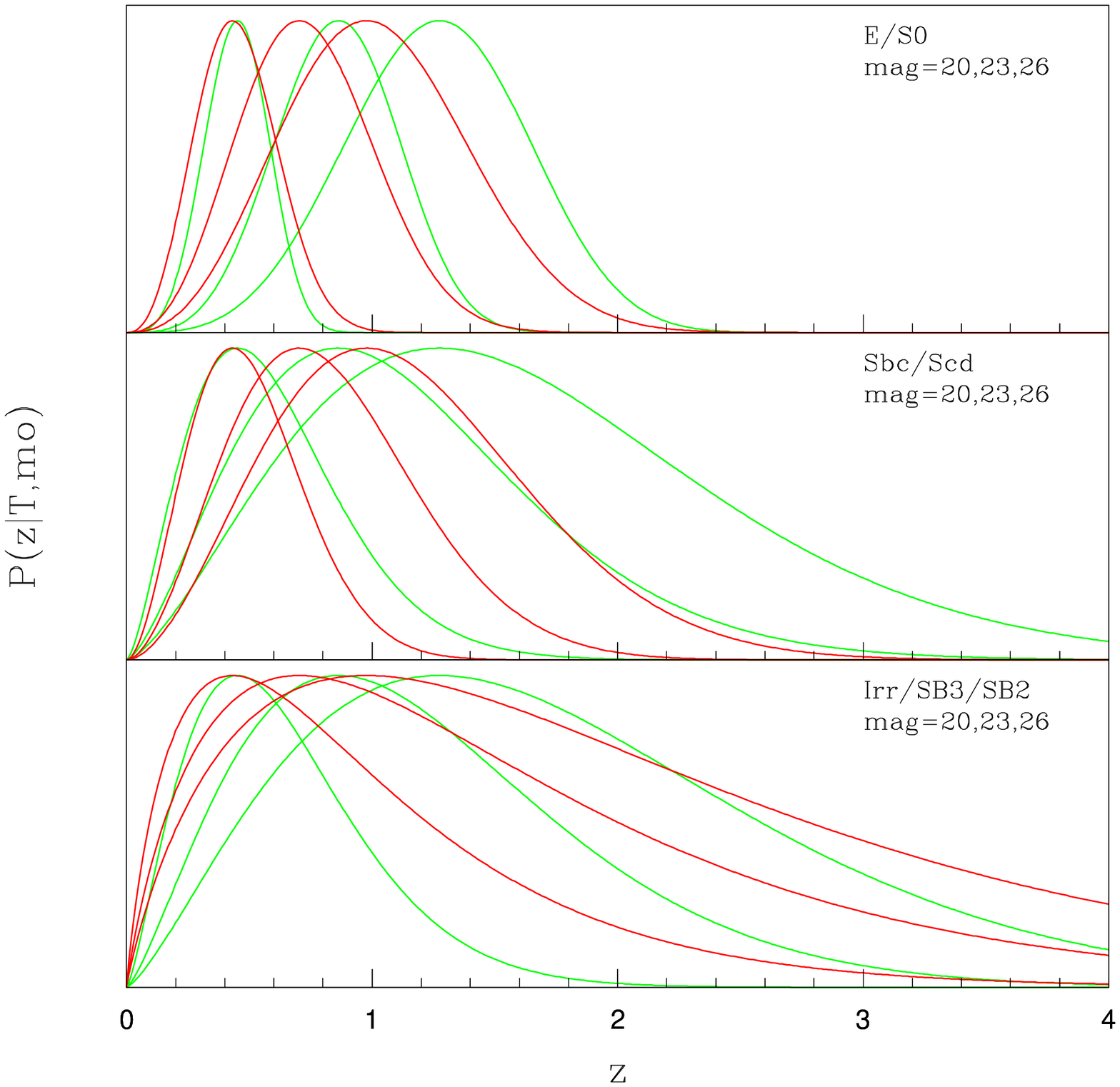}
\caption{Priors used to populate the simulations. Left: $P(T|m_0)$ is
the galaxy type fraction as a function of magnitude; Right:
$P(z|T,m_0)$ is the redshift distribution for galaxies of a given
spectral type and magnitude for $mag=20,23,26$. Throughout this paper
we use the priors indicated by the solid red lines (BPZ code). The
dashed red lines represent the priors in BPZ's paper (Benitez 2000),
while the green lines indicate the priors derived by Ilbert \etal
(2007).
\label{fig-priors}}
\end{figure}

We simulate a mix of ellipticals, spirals, irregulars and starburst
galaxies (specifically, E, Sbc, Scd, Irr, SB3, and SB2 templates)
following the priors for galaxy type fraction as a function of
magnitude, $P(T|m_0)$, and for the redshift distribution for galaxies
of a given spectral type and magnitude, $P(z|T,m_0)$, that are used in
\bpz 's Bayesian photometric redshift code. 
We found that in Table 1 of Benitez (2000), two numbers were
inadvertently switched, but the numbers were correct in the publicly
downloadable code.  Benitez (private communication) has confirmed that
the table should read $k_t=0.450$ for E/SO and $k_t=0.147$ for
Sbc/Scd.  Figure~\ref{fig-priors} shows in solid red lines the priors
used in this paper (same as \bpz 's code); in dashed red lines are the
priors quoted in BPZ's paper; and green lines represent Ilbert
\etal~(2007) priors.

\begin{figure}
\plotone{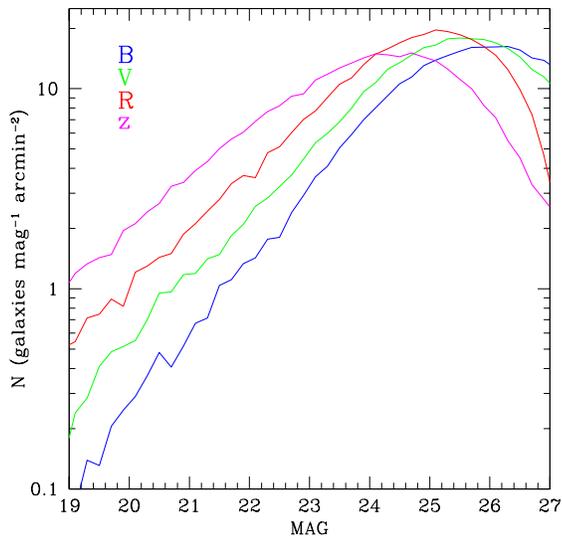}
\caption{DLS $N(mag_{iso})$ for BVRz.
\label{fig-nmagdls}}
\end{figure}

\begin{figure}
\plotone{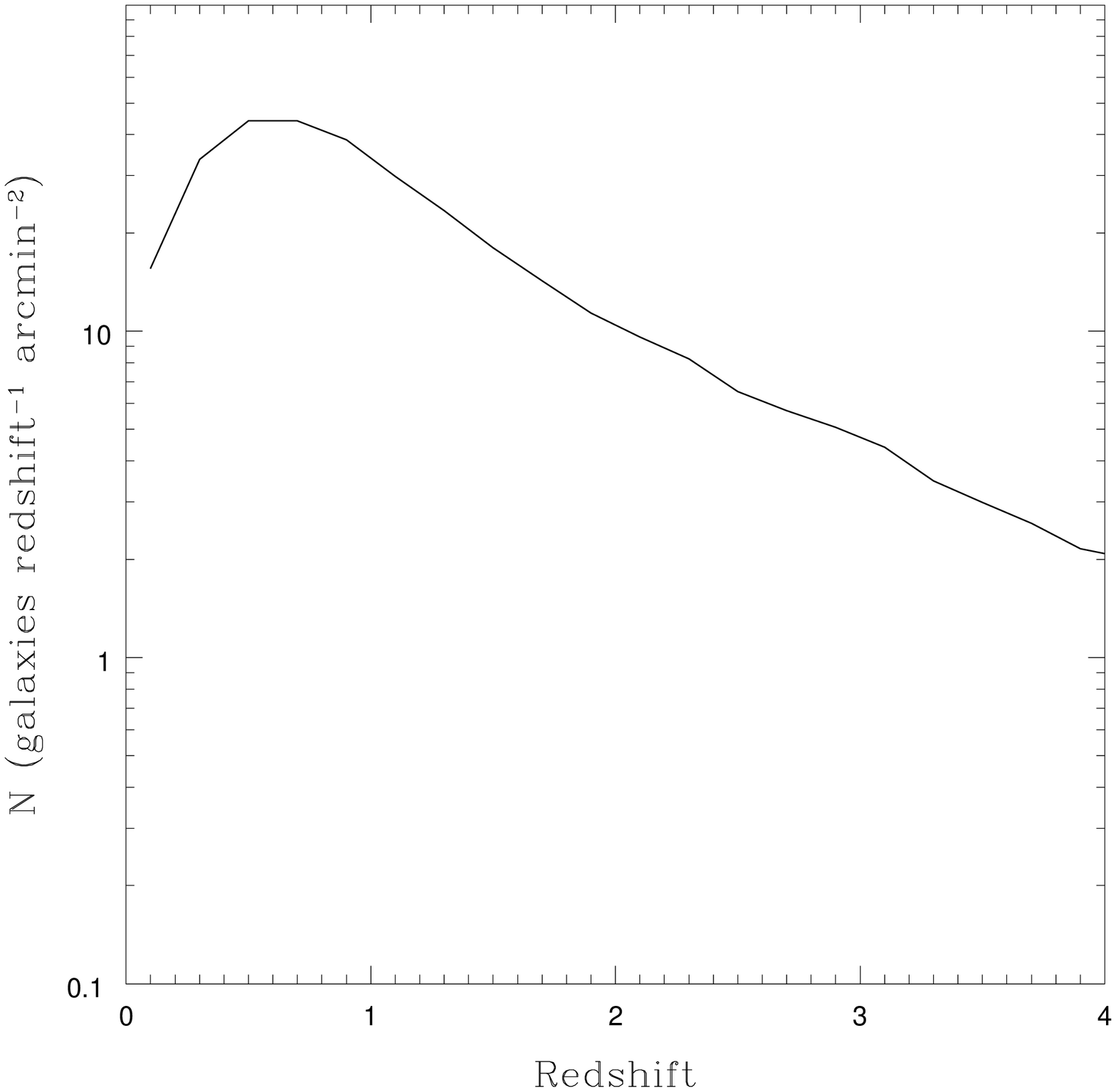}
\caption{$N(\zi)$ for simulations.
\label{fig-nzinput}}
\end{figure}

In order to have a realistic galaxy luminosity function, $N(mag)$, we
start our simulations from R-band magnitudes of 87260 objects detected
in one of our $\sim40^{\prime}\times40^{\prime}$ Deep Lens Survey
sub-fields (Wittman \etal 2002). The typical BVRz magnitude
distributions for the DLS are shown in Figure~\ref{fig-nmagdls}.  We
take this magnitude as the {\it true} ($\ne observed$) R-band
magnitude of a new object to be simulated. From the $P(T|m_0)$ prior
we select a \sed, and from $P(z|T,m_0)$ we choose a $\zi$ redshift for
the galaxy. The resulting ``true'' redshift distribution in the
simulations is shown in Figure~\ref{fig-nzinput}.  This distribution
has a larger tail to high redshift than usually found in the
literature (e.g. LeFevre et al 2005) and can be approximately described
as $z^{2} exp\big(-1(\frac{z}{0.05})^{0.54}\big)$. Magnitudes (with or
without noise) in any other photometric bands can then be computed. We
use \bpz~itself to compute synthetic colors, so there is no issue of
minor differences in the k-corrections, priors, etc.  We assume that
there are only six \sed `s of galaxies in the universe and make no
attempt to introduce template noise in these simulations.  We then
perform three sets of simulations in the BVRz filter set of the
DLS. In the first simulation (SIM1) we assume perfect, infinite
\sn~photometry. In the second set of simulations (SIM2) we
successively degrade the \sn~ of the photometry but maintain constant
the \sn~of all galaxies in all 4 bands (same magnitude error for all
galaxies in all 4 bands). In the third simulation (SIM3), we reproduce
the \sn~distribution and completeness of the DLS.

%-----------------------------------------------------------------%

\subsection{SIM1}\label{sim1}

\begin{figure}
\plotone{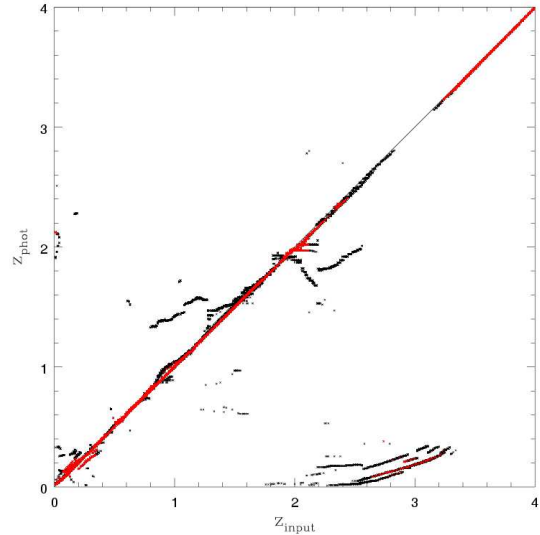}
\caption{$\zp-\zs$ scatter-plot for SIM1 (no photometry
  noise). Galaxies with $ODDS>0.9$ are in red. See first line of
  Table~\ref{tab:sim12} for statistics.
\label{fig-sim1}}
\end{figure}

\begin{figure}
\plotone{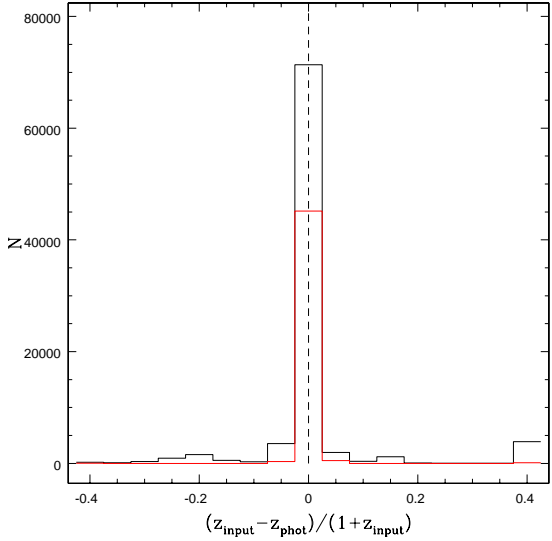}
\caption{Histogram of $\dz$ for the simulation in
  Figure~\ref{fig-sim1}. The distribution of galaxies with $ODDS>0.9$
  is shown in red. The outermost bins show the integrated counts of
  all objects with $|\dz|>0.4$.
\label{fig-sim1dz}}
\end{figure}

The first simulation (SIM1) has perfect photometry and represents the
best possible case. The $\zp-\zs$ scatter-plot for this simple
simulation is shown in Figure~\ref{fig-sim1}, and the distribution of
$\dz \equiv {\zs - \zp \over 1 + \zs}$ is shown in
Figure~\ref{fig-sim1dz}. Note that Figure~\ref{fig-sim1} contains
87260 objects, distributed in redshift according to
Figure~\ref{fig-nzinput}, and that the $\zp=\zs$ line is saturated
with objects. It is clear from Figure~\ref{fig-sim1dz} that the
majority of objects have $|\dz|\sim0.0$.  Table~\ref{tab:sim12}
indicates: (1) signal-to-noise of photometry (same in all bands); (2)
fraction of galaxies with $|\dz|<0.5$; (3) mean $\dz$ for galaxies
with $|\dz|<0.5$; (4) rms in $\dz$ for galaxies with $|\dz|<0.5$; (5)
fraction of objects with $ODDS>0.9$; (6) fraction of objects with
$ODDS>0.9$ and $|\dz|<0.5$; (7) mean $\dz$; and (8) rms in $\dz$ for
these galaxies.  

There are still catastrophic outliers, despite being the best possible
case in terms of noise, perfectly known templates, etc.  This is
because each galaxy is assigned a single $\zp$ based on the peak of
its PDF.  Consider a degeneracy such that the same colors come from
\sed~ A at $z_1$ or \sed~ B at $z_2$.  In the absence of priors, this
would result in a PDF with two equal peaks.  Now add priors encoding
our astrophysical knowledge, such as that an apparently bright galaxies
are likely to be at low redshift, or that ellipticals are rare at high
redshift.  This usually helps select the correct peak, but sometimes
it will select the wrong peak because unlikely events do happen: some
high-redshift galaxies are bright, or are ellipticals.  As noted
above, this ignores the full PDF, which may be broad or multi-modal in
a way that is consistent with the true redshift.  As our purpose is
only to establish SIM1 as a baseline for investigating the impact of
photometric S/N, we do not pursue this here.

%-----------------------------------------------------------------%

\subsection{SIM2}\label{sim2}

\begin{figure}
\plotone{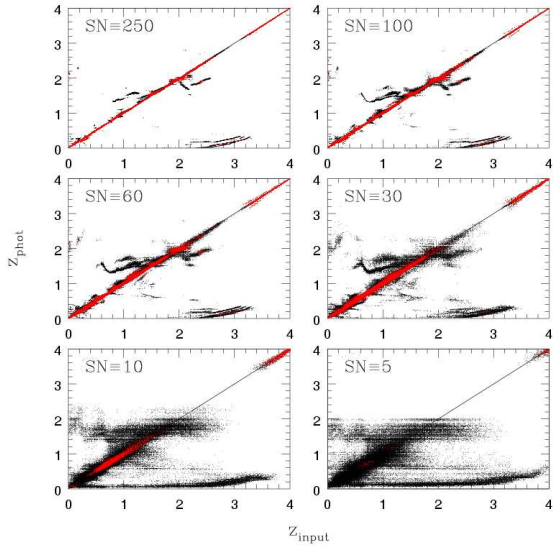}
\caption{$\zp-\zs$ scatter-plot for simulations with 
realistic magnitude and redshift distributions, but uniform and
progressively greater photometry noise (SIM2). In each panel all
galaxies have the same \sn~ in BVRz. From top-left to bottom-right:
SN=250, 100, 60, 30, 10, 5. See Table ~\ref{tab:sim12} for statistics.
\label{fig-sim2}}
\end{figure}

\begin{figure}
\plotone{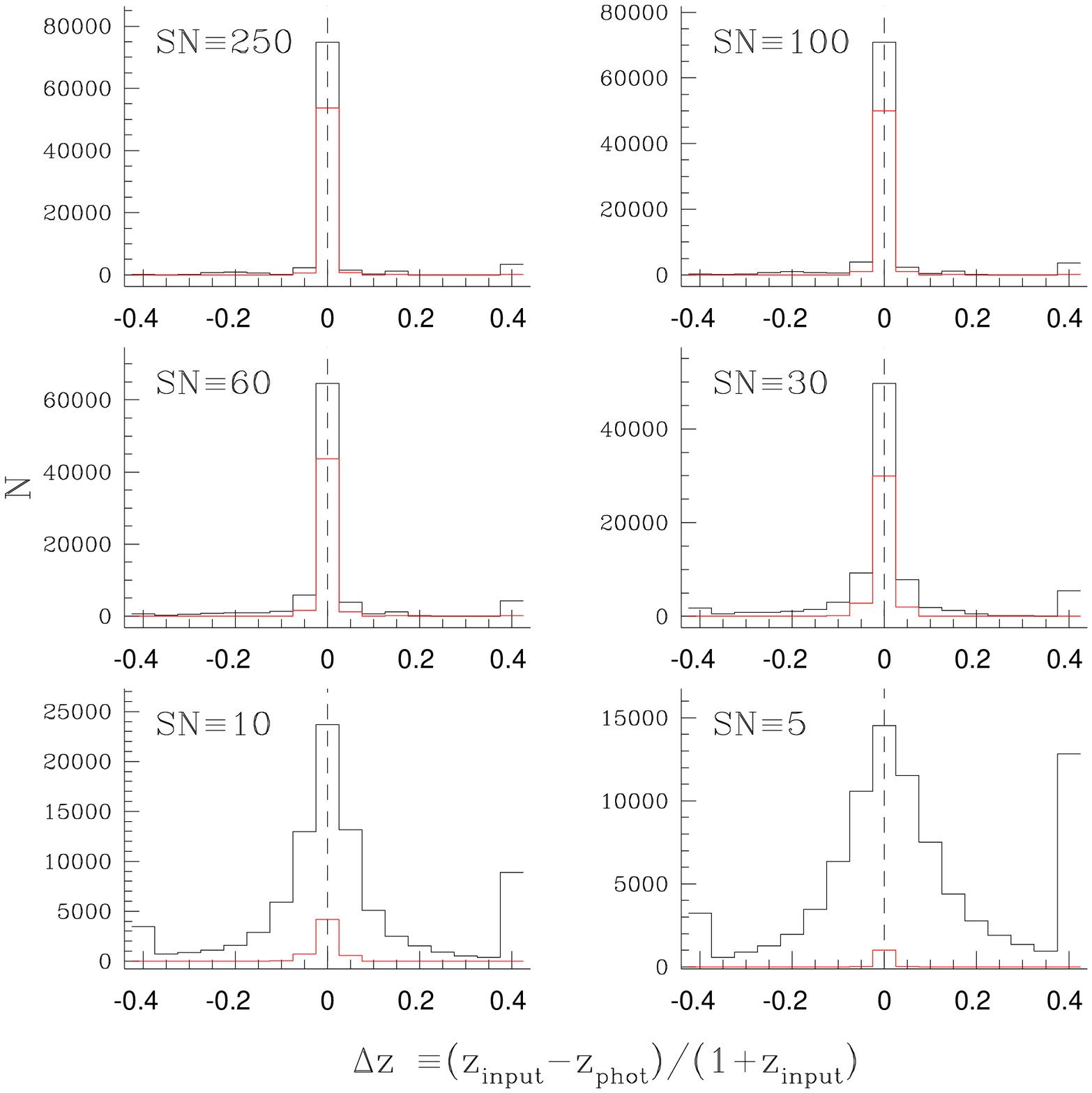}
\caption{Histogram of $\dz$ for objects shown in Figure~\ref{fig-sim2} (SIM2).
\label{fig-sim2dz}}
\end{figure}

\begin{figure}
\plottwo{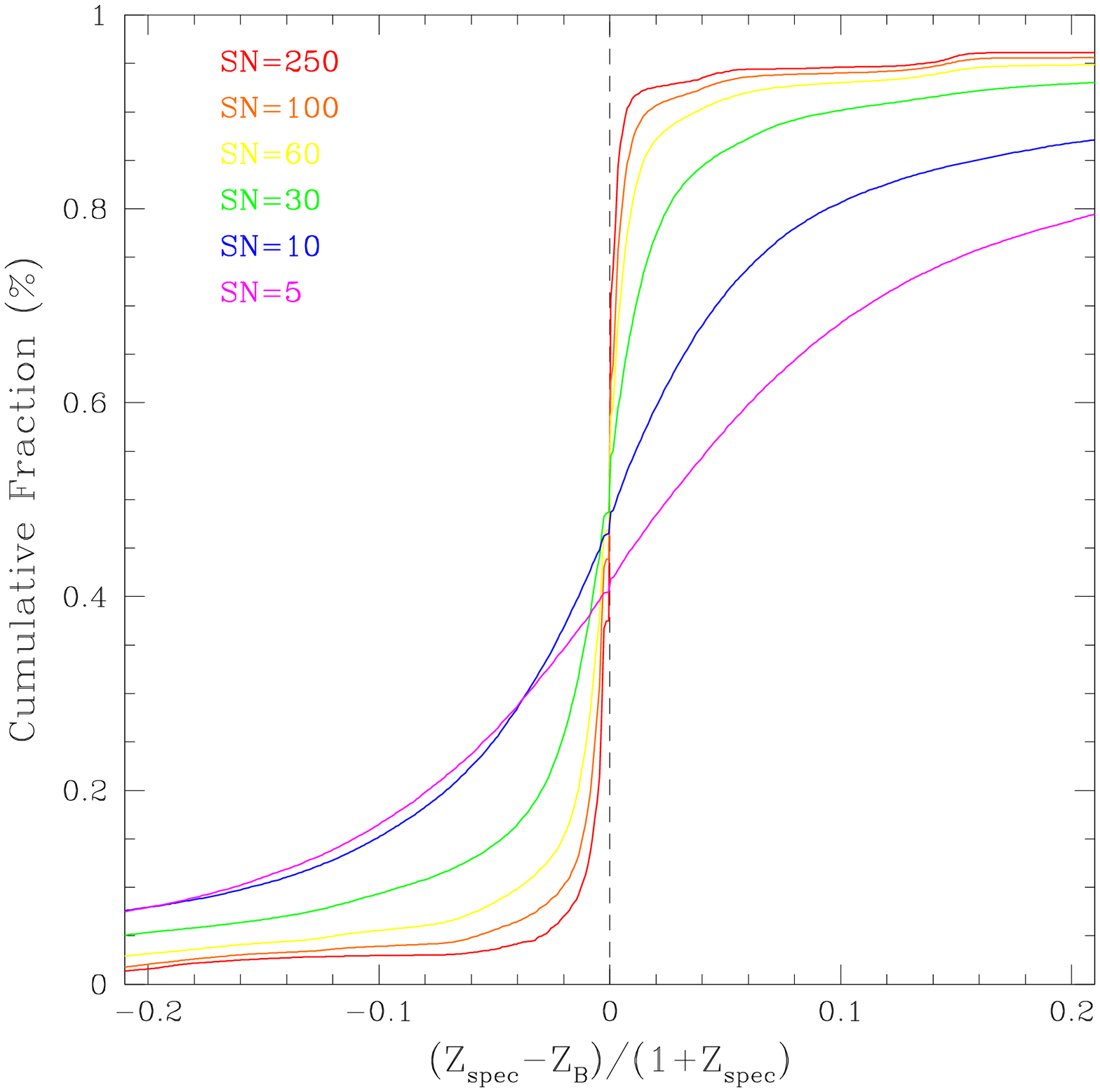}
        {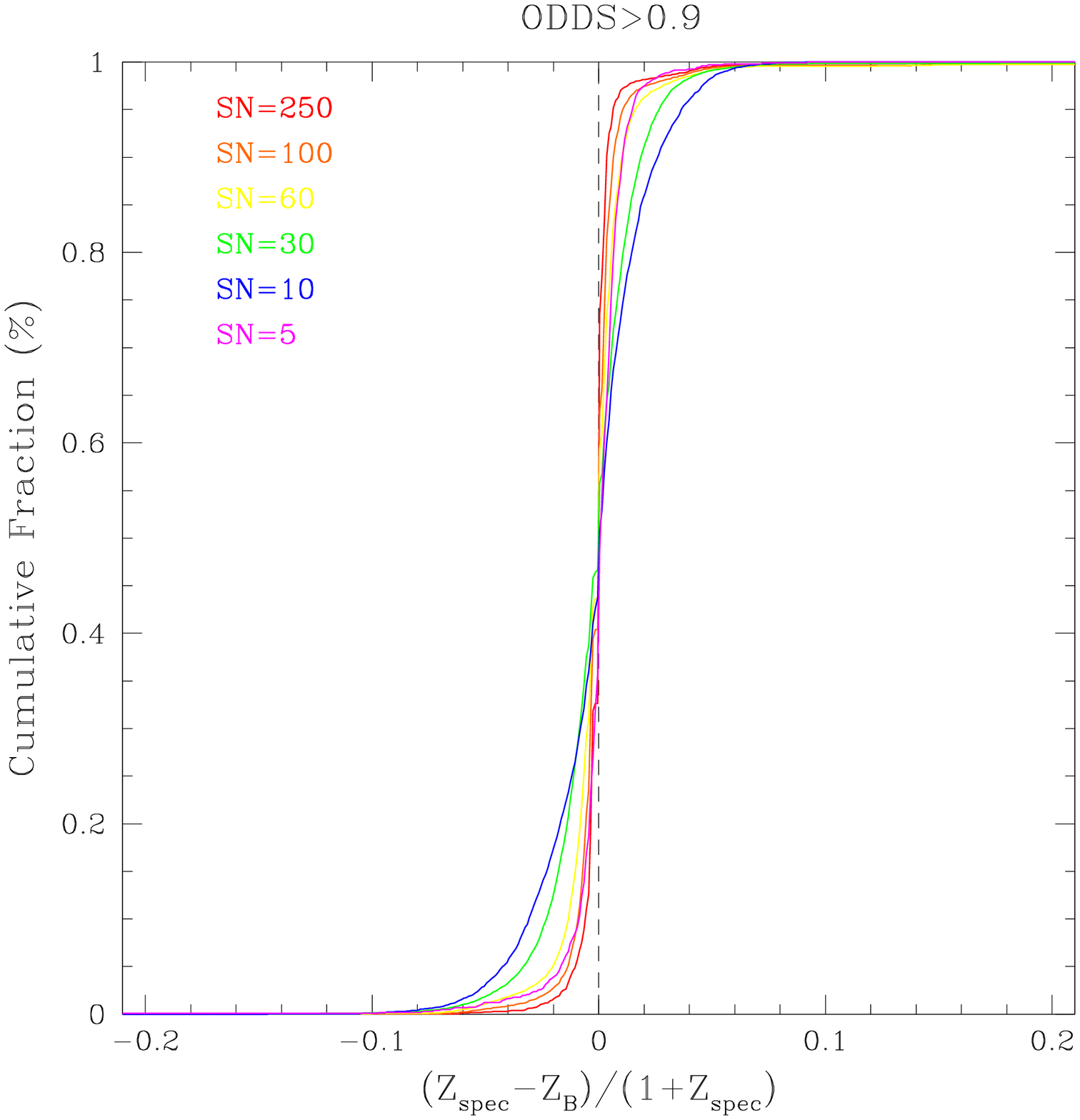}
\caption{Cumulative fraction of objects with $\dz$ smaller than a
  given value. Red line indicates the simulation in which all galaxies
  have been set to have $SN=250$ in all BVRz; orange indicates a
  simulation with $SN=100$; and so on. Right panel shows all galaxies,
  and left panel shows galaxies with $ODDS>0.9$. Note that only 6.4\%
  and 1.2\% respectively of objects with $SN=10,5$ have $ODDS>0.9$.
\label{fig-sim2frac}}
\end{figure}

In the second set of simulations (SIM2) we degrade the initially
perfect photometry in SIM1 successively to \sn~ of 250 ($R\sim20.5^m$
in the DLS, and the magnitude limit of the spectroscopic sample
presented in Section \ref{data}), 100, 60, 30, 10 and 5, and repeat
the analysis at each step. In these unrealistic simulations all
galaxies have the same photometric \sn~ in all bands. The
scatter-plots are shown in Figure~\ref{fig-sim2}, and $\dz$
distributions are shown in Figure~\ref{fig-sim2dz}. We also present
the cumulative fraction of objects with $\dz$ smaller than a given
value, as a function of $\dz$ (Figure~\ref{fig-sim2frac}). This plot
has several advantages.  First, multiple simulations can be
over-plotted without obscuration.  Second, the asymmetry in the
distribution of $\dz$ is easily read off by looking at the fraction
with $\dz<0$ (dashed vertical line).  Third, the fraction of outliers
can also be directly read off the plot at any $\dz$. The left panel of
Figure~\ref{fig-sim2frac} shows the cumulative fraction for all
objects, while the right panel shows $ODDS>0.9$ galaxies. The number
of galaxies in the right panel is smaller than the number in the left
(see Table~\ref{tab:sim12}) but the accuracy of photo-zs is clearly
better.

Because all realizations of SIM2 have the redshift distribution shown
in Figure~\ref{fig-nzinput}, even if all galaxies have colors measured
at very high \sn~, some objects will have degenerate colors and the
sample will contain some fraction of catastrophic
outliers. Spectroscopic samples typically have a much lower mean
redshift than these simulations, so catastrophic outliers are likely to
be underrepresented in direct $\zp-\zs$ comparisons, if the full
photometric sample is very deep.

Table~\ref{tab:sim12} presents the statistics for the SIM2 objects
shown in Figures~\ref{fig-sim2}, ~\ref{fig-sim2dz} and
~\ref{fig-sim2frac}.  Clearly, the precision of photometric redshifts
is a strong function of photometric \sn.  BPZ's ODDS parameter is very
effective at removing outliers, and almost 100\% of the objects with
$ODDS>0.9$ have $|\dz|<0.1$ regardless of \sn~(right panel in
Figure~\ref{fig-sim2frac}). However, the fraction of objects with
$ODDS>0.9$ decreases dramatically with decreasing \sn.

Performance is, counter-intuitively, slightly worse for the infinite
S/N galaxies in SIM1 than for the high S/N galaxies in SIM2.  This is
because the priors have too much power when there is no noise in color
space, and is not of concern in more realistic situations.

%-----------------------------------------------------------------%

\subsection{SIM3}\label{sim3}

The third simulation has the same \sn~ distribution and completeness as
the DLS data. Again, the priors used assure that the galaxy type
mixture and redshift distribution should be close to the real
universe.  The idea is to measure how well we can recover true $\zi$
redshifts for a realistic photometric data set. This simulation is
still optimistic because no template noise is added---we derive colors
from the same six templates used in the determination of photometric
redshifts. The effect of template noise will be presented in the real
data analysis in Section \ref{data}.

\begin{figure}
\plotone{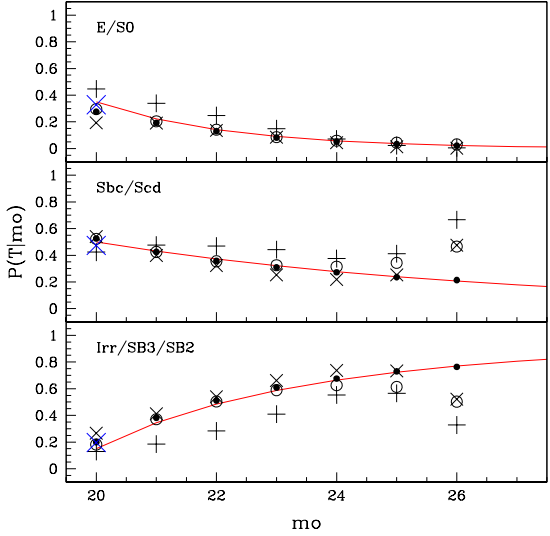}
\caption{Galaxy type fraction as a function of magnitude,
  $P(T|m_0)$. The solid red lines indicate BPZ's priors used in the
  simulations of Section~\ref{simulations} and shown previously in
  Figure~\ref{fig-priors}. The ``$+$'' and ``$\times$'' symbols
  indicate fraction of galaxies classified by BPZ as E, Sbc/Scd, or
  Im/SB3/SB2 in two DLS fields of $40^{\prime}\times40^{\prime}$
  each. The ``$+$'' field, with higher fraction of ellipticals,
  contains the galaxy cluster Abell 781, while the ``$\times$'' 
  represents a more typical ``blank'' field. The simulation input
  distribution is indicated by solid circles, which by definition
  agree with the red line, while the open circles indicate the BPZ
  type classification of these objects. The blue ``$\times$''
  represents the SHeLS spectroscopic sample.
\label{fig-priors2}}
\end{figure}

As a sanity check we compare the BVz magnitude distributions of our
SIM3 simulation with the observed $N(mag)$ and find good
agreement. The R magnitude distribution is by definition the same
within the added photometric noise. We also compare the distribution
of BPZ galaxy types in DLS fields with the one derived from the SIM3
simulation and find very good agreement.  Figure~\ref{fig-priors2}
shows the galaxy type fraction as a function of magnitude for two
$40^{\prime}\times40^{\prime}$ DLS fields. The field with the higher
fraction of ellipticals contains the richness class 2 galaxy cluster
Abell 781 at $z=0.298$ (``$+$''), and the other is a more typical
``blank'' field (``$\times$''). The simulation input distribution is
indicated by solid circles, which by definition agree with the red
line, and the output BPZ types are indicated by open circles. SIM3 and
data show the same magnitude dependence.

\begin{figure}
\plotone{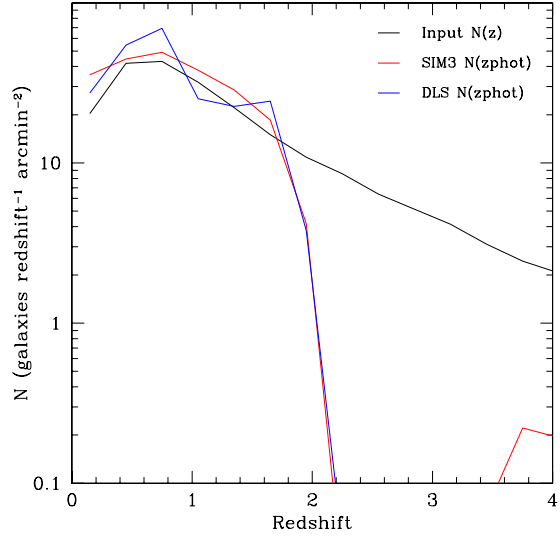}
\caption{Photometric redshift distributions for the DLS (blue) and SIM3 (red).
  The input $N(z)$ for the simulations is shown in black.
\label{fig-dlsnz}}
\end{figure}

A third sanity check is a comparison between the redshift distribution
derived in SIM3 and $N(z)$ for the entire DLS survey.
Figure~\ref{fig-dlsnz} shows both distributions and also the input
redshift distribution used in the simulations (same as shown in
Figure~\ref{fig-nzinput}). The agreement is pretty good. The mean
density of galaxies with photometric redshifts of any quality is
$47/arcmin^2$ and $11\%$ of those objects have $ODDS>0.9$.

\begin{figure}
\plotone{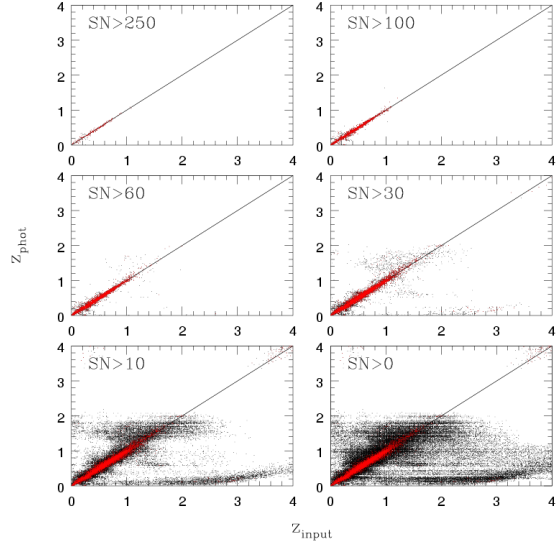}
\caption{$\zp-\zs$ scatter-plot for simulations with realistic
 magnitude, redshift, and S/N distributions (SIM3).  Upper left panel
 shows all galaxies with very high \sn~ (small number density and
 mostly at low redshift) while lower right panel includes all galaxies
 in the simulation. From upper left to lower right: (1) galaxies with
 $SN>250$; (2) $SN>100$; (3) $SN>60$; (4) $SN>30$; (5) $SN>10$; and
 (6) $SN=All$.
\label{fig-sim3}}
\end{figure}

\begin{figure}
\plotone{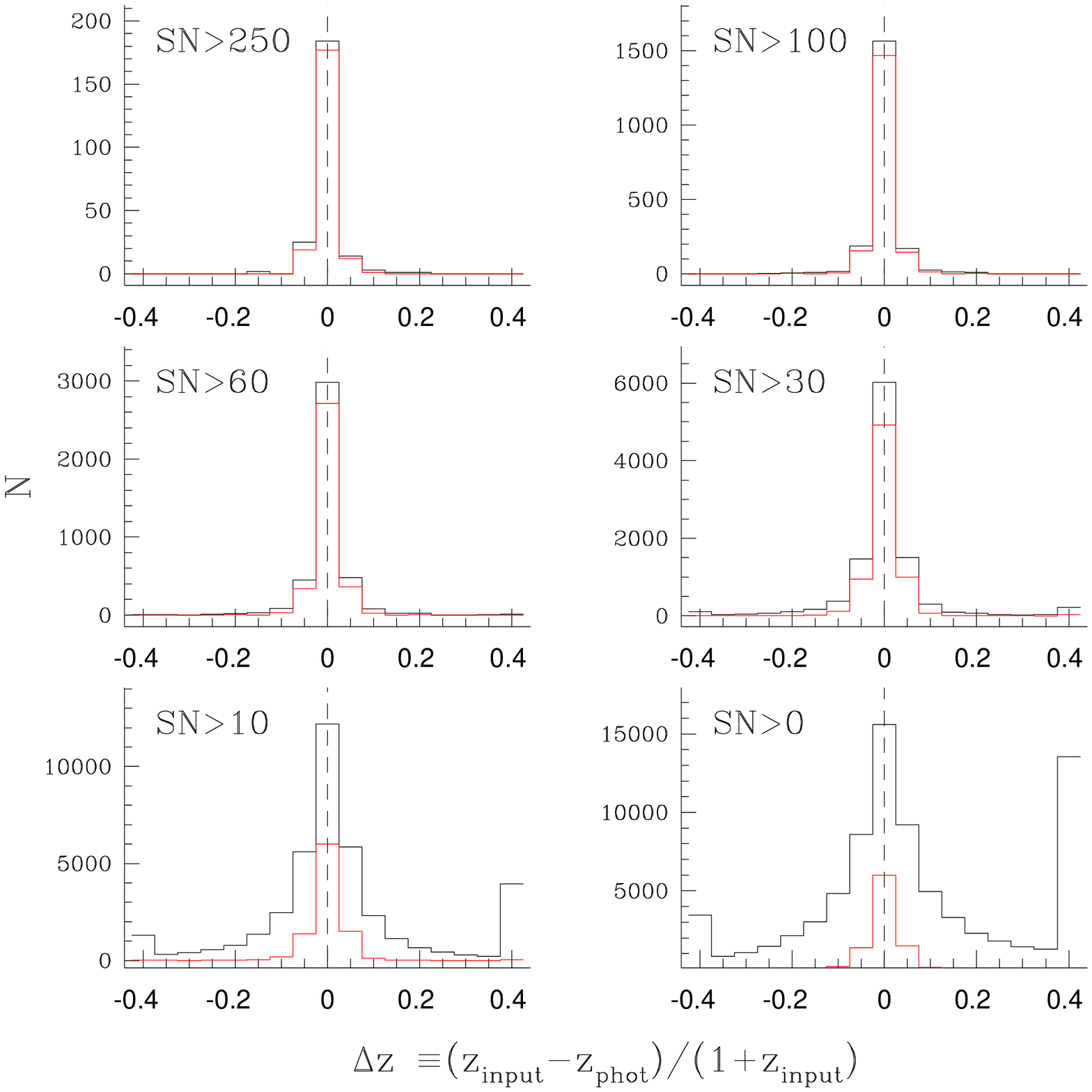}
\caption{Histogram of $\dz$ for objects shown in Figure~\ref{fig-sim3}.
\label{fig-sim3dz}}
\end{figure}

\begin{figure}
\plottwo{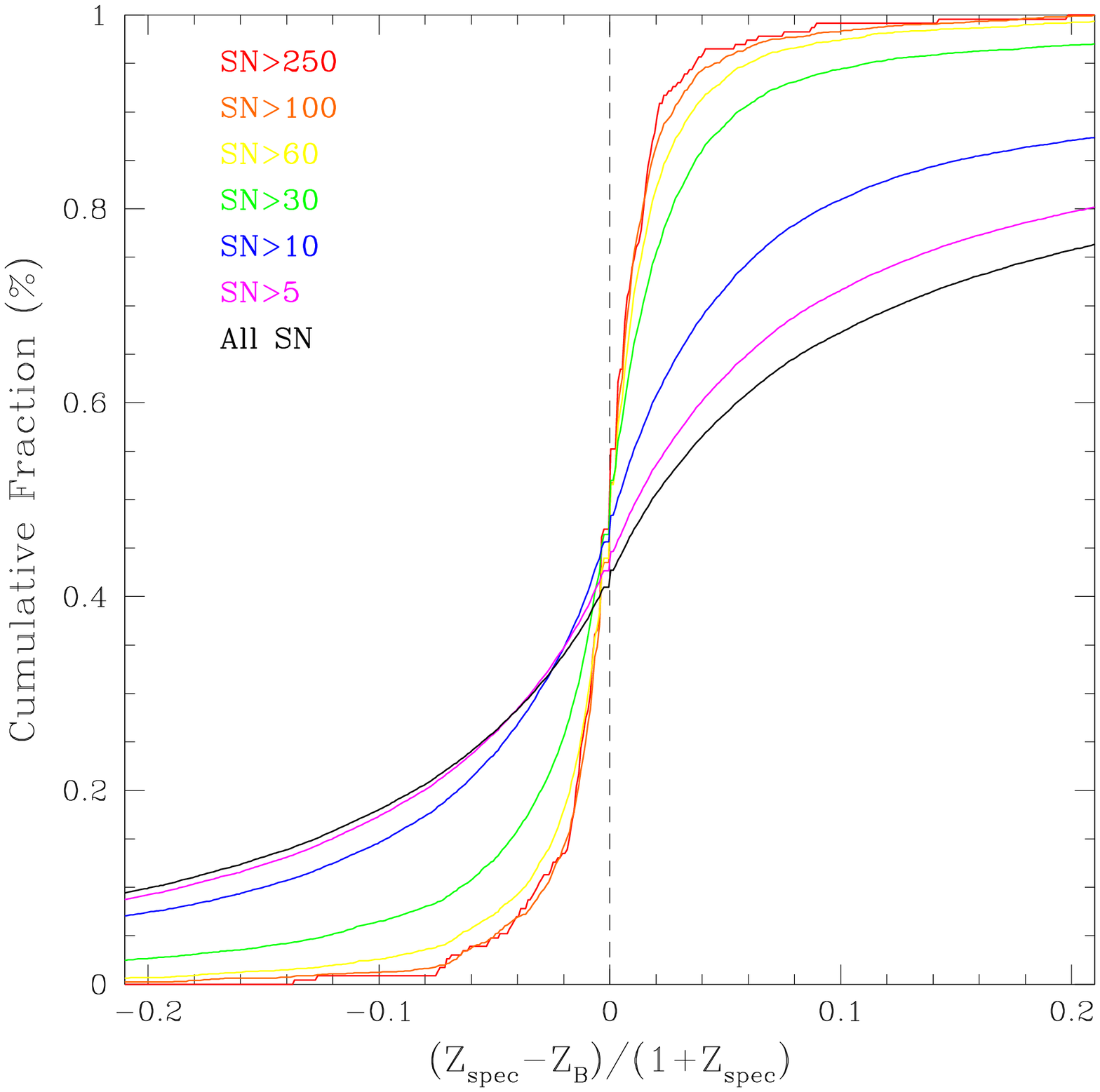}
        {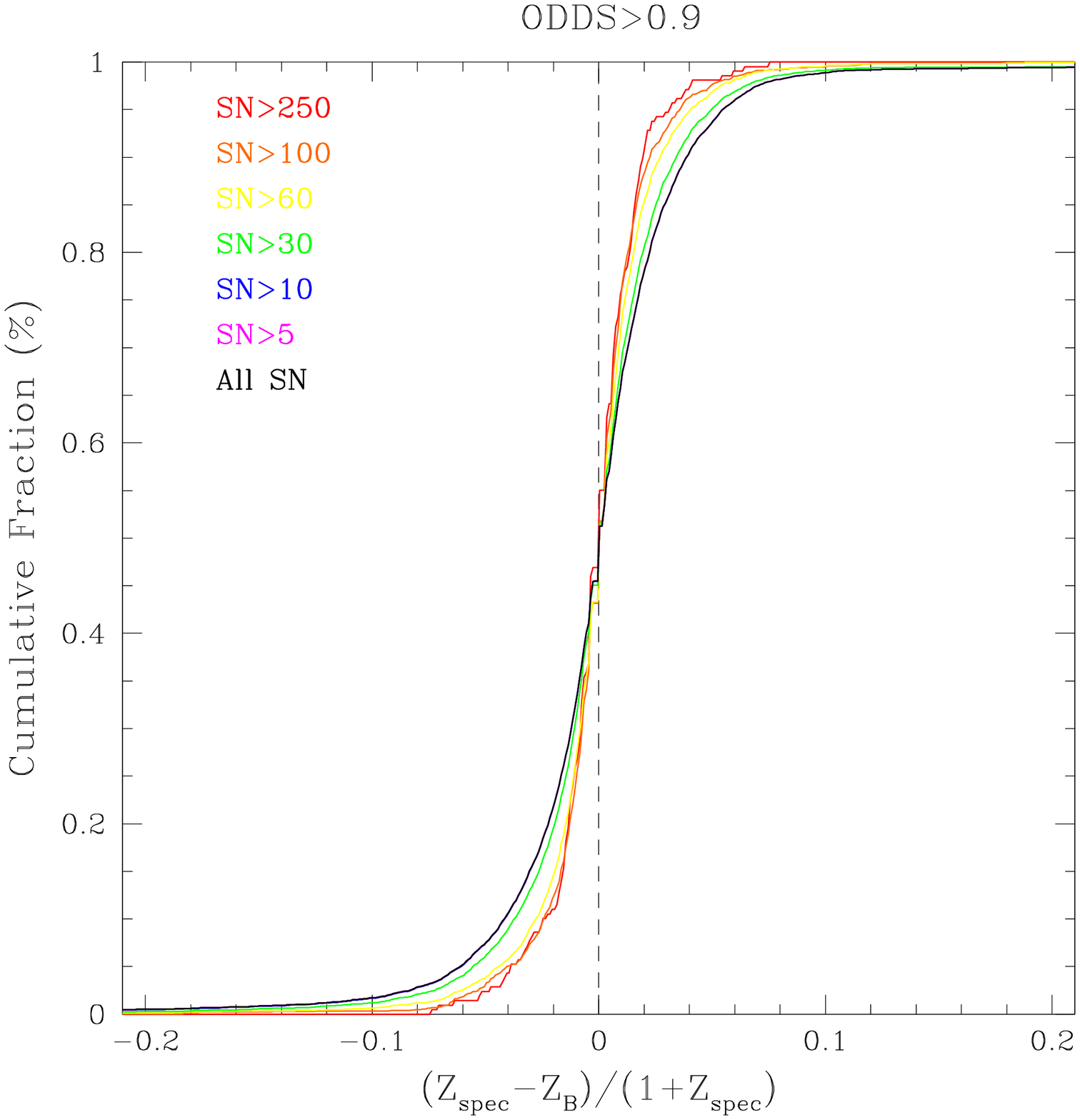}
\caption{Cumulative fraction of objects with $\dz$ smaller than a
  given value for simulations with realistic magnitude and redshift
  distributions (SIM3). Red line shows the cumulative fraction for all
  objects with $SN(R)>250$; orange shows the fraction for all objects
  with $SN(R)>100$ (including those with $SN(R)>250$); and so on.
  Right panel shows all galaxies, and left panel shows galaxies with
  $ODDS>0.9$.
\label{fig-sim3frac}}
\end{figure}

The photometric redshift performance on SIM3 is shown on
Figures~\ref{fig-sim3}, ~\ref{fig-sim3dz} and~\ref{fig-sim3frac}, just
as in Figures~\ref{fig-sim2}, ~\ref{fig-sim2dz} and~\ref{fig-sim2frac}
for SIM2.  The summary statistics for SIM3 are presented in
Table~\ref{tab:sim3}.  As in SIM2, the precision of photometric
redshifts is a strong function of \sn, and ODDS does a good job of
cleaning up, at the cost of losing many low \sn~galaxies.

There are two notable differences with SIM2.  First, in SIM3, there is
a realistically strong correlation between high S/N and bright
magnitudes.  A bright magnitude implies a strong prior (most bright
galaxies are at low redshift), whereas a faint galaxy has a weak prior
(it could be at any redshift).  The high S/N galaxies in SIM2 were
(artificially) at all magnitudes, and therefore had generally looser
priors.  Therefore, the highest S/N galaxies in SIM3 do better than
those in SIM2.  We can see the effect of the tight priors directly by
comparing the $S/N=250$ line of Table ~\ref{tab:sim12}
($\sigma_{\dz}=0.042$ after clipping 4\% which had $|\dz|>0.5$) with
that of Table~\ref{tab:sim3} ($\sigma_{\dz}=0.031$ with no need to
clip any outliers).  This difference vanishes when low S/N galaxies
from SIM3 are included.

In fact, the $S/N=5$ galaxies in SIM2 outperform the $S/N>5$ galaxies
in SIM3, despite the latter cut being only a lower bound.  This is due
to the second salient difference between SIM2 and SIM3: A given S/N in
SIM2 describes {\it each} galaxy in {\it each} band.  In SIM3, the S/N
varies with filter in a realistic way, and the cut applies to R band.
Most galaxies will have lower S/N in other bands. For $S/N=30$ in R,
the median $S/N$ in B, V, and z over the whole sample is 10, 18, and
10 respectively.

%The third difference between SIM2 and SIM3 is that in SIM2,
%$\sigma_{\dz}$ is extremely small ($\sim0.01$) after applying the ODDS
%cut.  This is artificially good, because photometric calibration
%errors were not introduced and marginalized over as with SIM3.  SIM3
%represents our best estimate of performance in the real DLS survey.

What S/N is required for good photometric redshift performance?
First, consider performance without any ODDS cut.  At each step in
Table~\ref{tab:sim3} from $S/N>100$ to $S/N>10$, there is a 30--50\%
increase in $\sigma_{\dz}$, so there is no natural breakpoint.
$\sigma_{\dz}$ appears to stop this dramatic growth when stepping down
from $S/N>10$ to $S/N>5$, but this is likely an artifact of clipping
at $|\dz|>0.5$, which is roughly three times the clipped rms at that
point. Even at $S/N>10$, $\sigma_{\dz}$ may be artificially low due to
clipping, as more than 10\% of galaxies were clipped.  Most survey
users would find the precision offered by the $S/N>30$ cut acceptable,
but the $S/N>10$ cut unacceptable.  If we set $\sigma_{\dz}=0.1$ as
the limit of acceptability, we find an S/N cut at 17 is required.

Now consider using the ODDS cut at 0.9.  $\sigma_{\dz}$ is always 0.04
or less, regardless of S/N. We suspect that for a given
$\sigma_{\dz}$, the ODDS cut will provide more galaxies than the S/N
cut, because ODDS responds to the properties of the color space as
well as to S/N.  For example, high-precision S/N is not required if
the galaxy is in a distinctive region of color space.  In addition,
ODDS can take proper account of different S/N in different bands,
which a simple S/N cut in R does not.  We investigate this by finding
the ODDS cut which yields the same $\sigma_{\dz}$ as the $S/N>30$ cut
(0.076).  We find that $ODDS>0.57$ is required, which yields 30\% of
all detected galaxies, vs. the 13\% yielded by the S/N cut.

We repeat this procedure for $\sigma_{\dz}=0.1$.  The required ODDS
cut is $>0.40$, yielding 45\% of all detected galaxies, while the
required S/N cut at 17 yields only 26\% of detected galaxies.

\begin{figure}
\plottwo{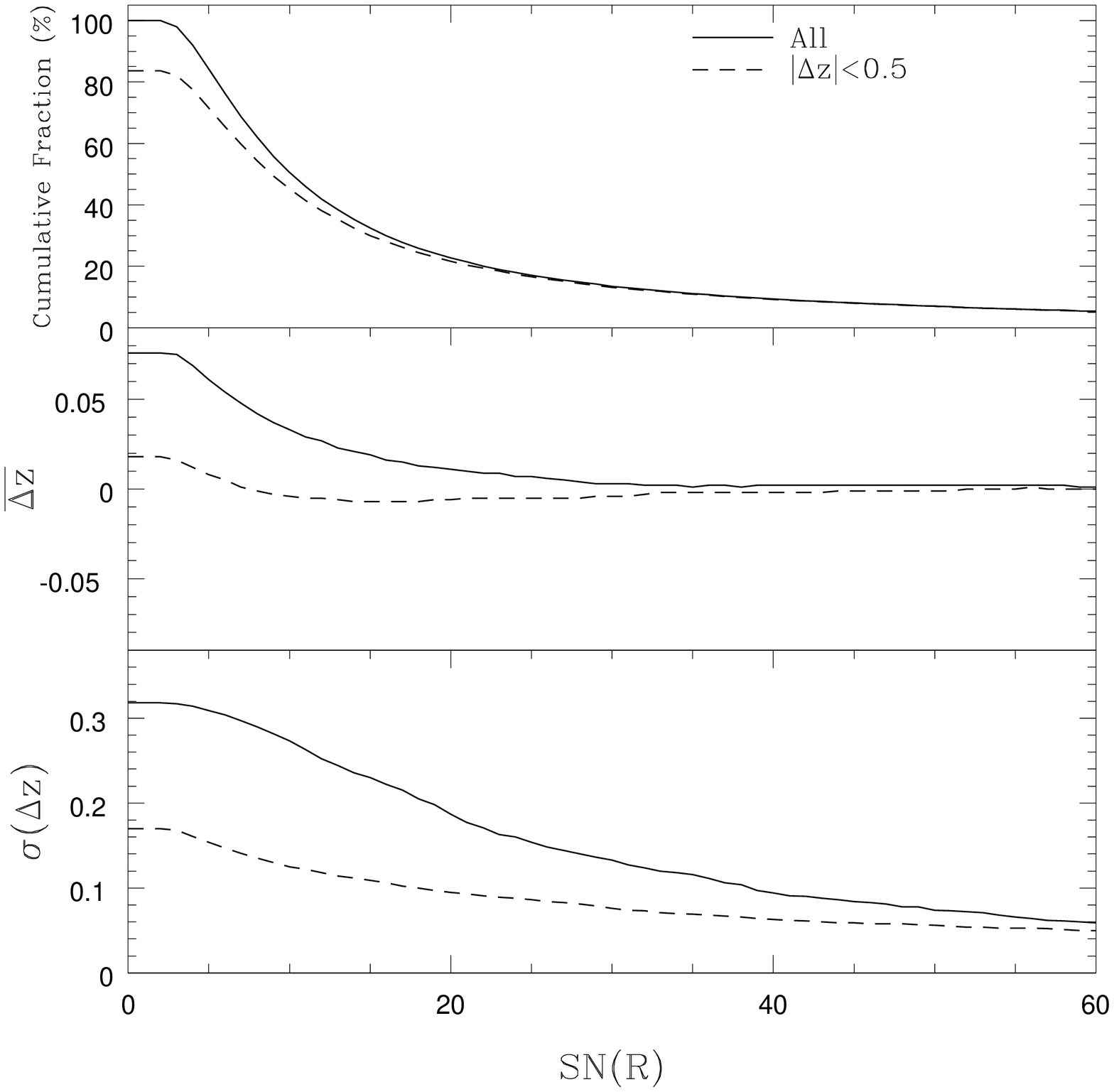}
        {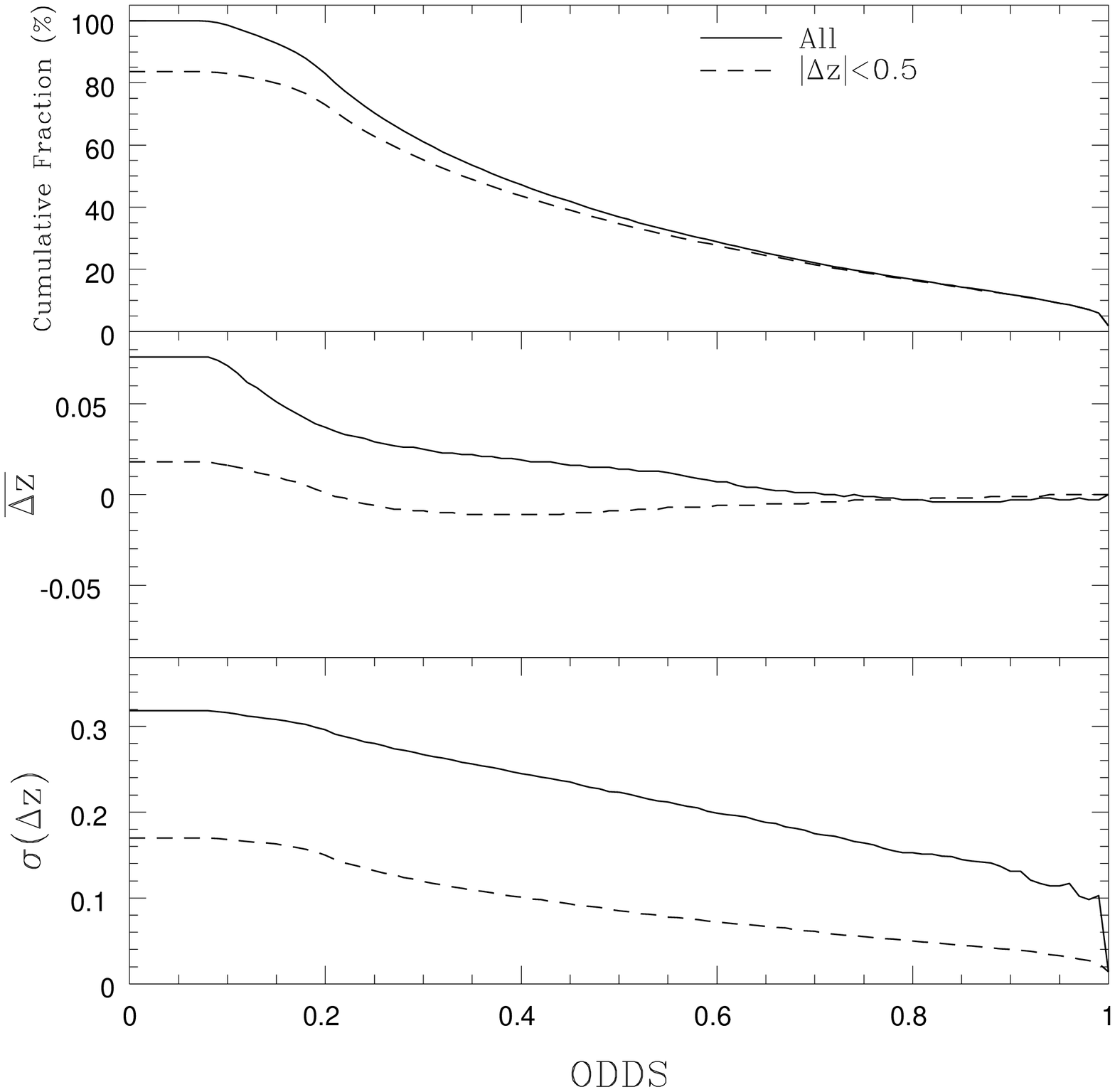}
\caption{{\bf Left:} Cumulative fraction of objects with \sn~greater than a 
  given value, mean $\dz$, and $\sigma_{\dz}$. {\bf Right:} same as
  left panels for objects with $ODDS$ greater than a given
  value. Solid line indicates all objects and dashed lines shows
  $|\dz|<0.5$.
\label{fig-sim3snodds}}
\end{figure}

These fractions can all be read off
Figure~\ref{fig-sim3snodds} which summarizes the results from
SIM3. The three left panels in Figure~\ref{fig-sim3snodds} show: (1)
the cumulative fraction of objects with \sn~greater than a given
value; (2) mean $\dz$; and (3) $\sigma_{\dz}$ for these objects. The
three right panels are the same but for a cut in $ODDS$.

In short, we recommend an ODDS cut.  We recognize that an ODDS cut is
not easy to incorporate into survey forecasts of the number of usable
galaxies.  Detailed simulations for a given filter set and depth as a
function of wavelength must be performed.  However, we hope that the
above numbers can serve as a rough guide for translation between
photometric redshift precision, S/N threshold, and number of usable
galaxies.

%-----------------------------------------------------------------%
%-----------------------------------------------------------------%

\section{Data}\label{data}

We take photometric data from the \dls~BVRz full-depth images in
fields with spectroscopic redshifts from the {\it Smithsonian
Hectospec Lensing Survey} (SHeLS, Geller \etal\ 2005), and from the
{\it Caltech Faint Galaxy Redshift Survey} (CFGRS, Cohen \etal~1999)
surveys. Here, by definition, template noise is present.  In Sections
\ref{shels} and \ref{cfgrs} we present the spectroscopic data and the
photometric redshift accuracy for these two samples, but before that
we present our methodology for color measurement (Section
\ref{colors}), and template optimization (Section \ref{tempopt}).

%-----------------------------------------------------------------%

\subsection{Measuring Colors}\label{colors}

\begin{figure}
\plottwo{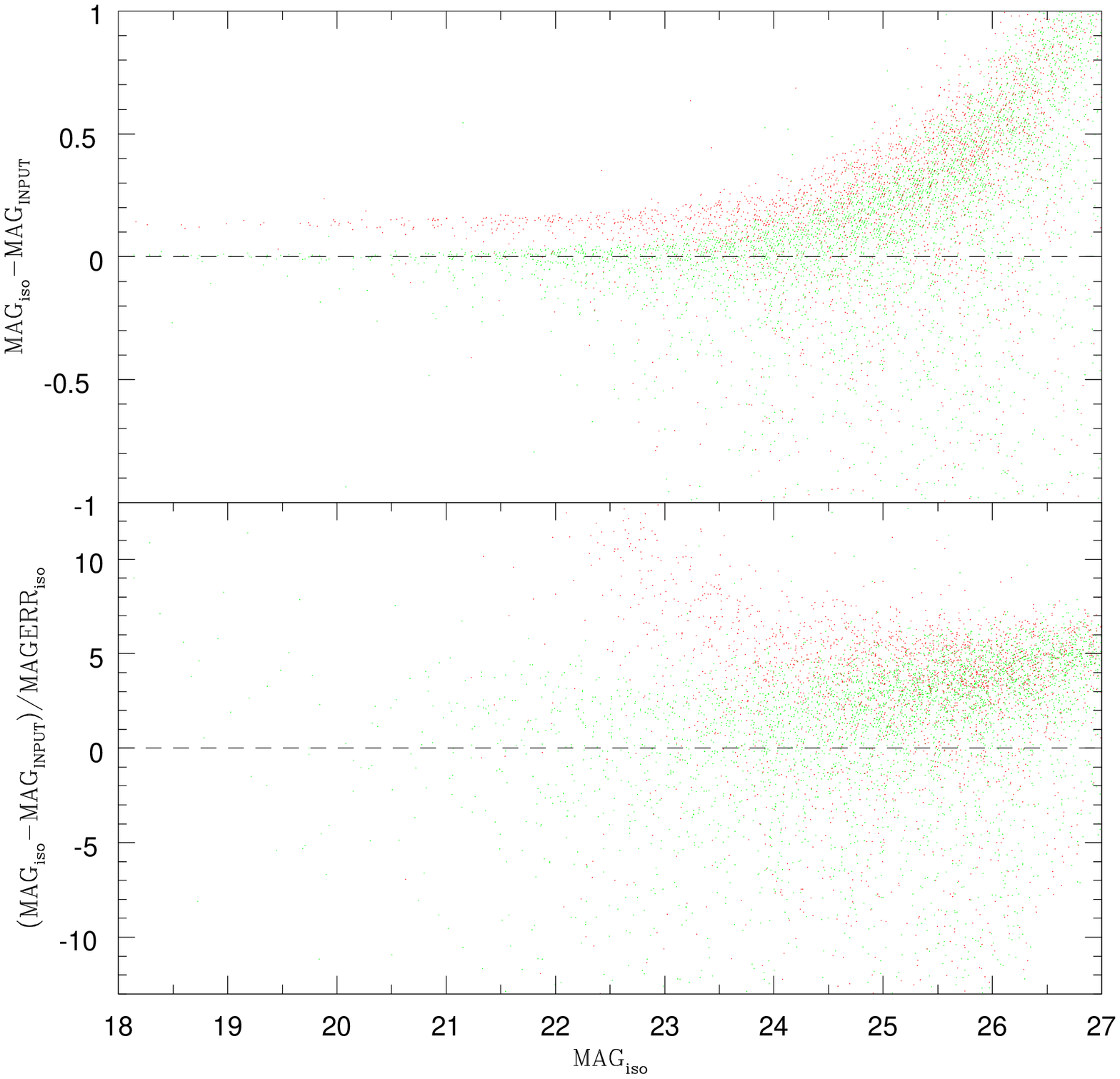}
        {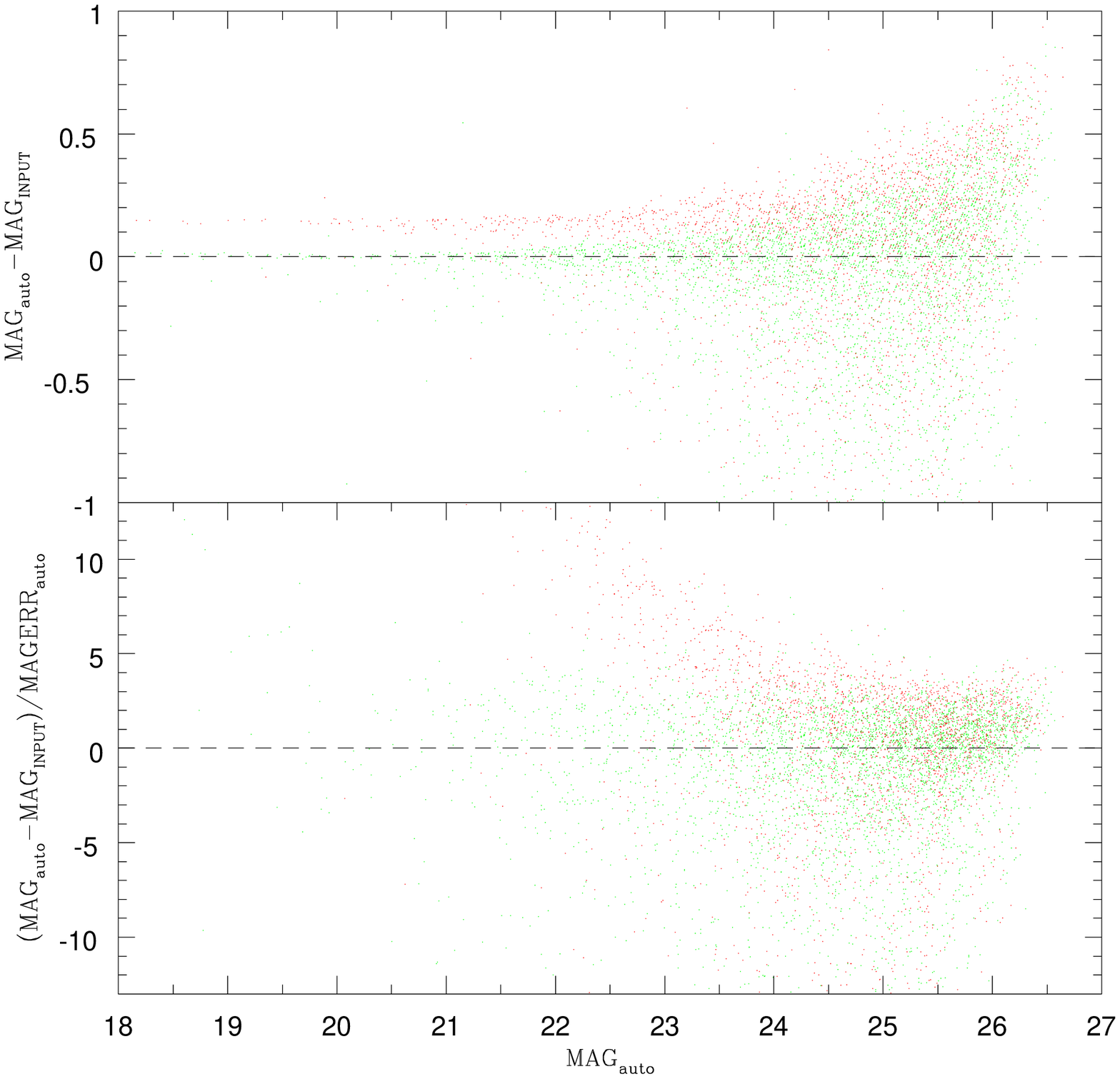}
\caption{Magnitude errors of synthetic De Vaucouleurs (red) and
  exponential disk (green) galaxies added to DLS R-band images. Left
  panels show the SExtractor $MAG_{ISO}$ results while right panels
  indicate comparisons to SExtractor $MAG_{AUTO}$.  Top: $\Delta{MAG}
  \equiv MAG-MAG_{input}$ {\it vs.} $MAG$. Bottom:
  $\Delta{MAG}/MAGerr$ {\it vs.} $MAG$.
\label{fig-magRmagerr}}
\end{figure}

We performed simulations to determine the best photometry method in
the face of different point-spread function (PSF) sizes in the
different filters.  We added galaxies with De Vaucouleurs (elliptical)
and exponential disk (spirals) light profiles to the \dls~BVRz data
using standard IRAF-Artdata routines, ran SExtractor (Bertin \&
Arnouts 1996) and measured colors with many different types of
magnitudes.  Figure~\ref{fig-magRmagerr} shows the results for
galaxies added to the R images. The B, V and z results are
qualitatively the same, but because of differences in \sn~and PSF
there is a shift in the magnitude axis, and slightly different
scatter. The left panels show the results using $MAG_{iso}$ and right
panels show $MAG_{auto}$.  The top panels show the difference between
measured $MAG$ and input $MAG_{input}$. De Vaucoleurs galaxies are
measured to be $\sim0.15^m$ fainter than their true magnitudes both by
$MAG_{iso}$ and $MAG_{auto}$. The bottom panels show the distribution
of $(MAG-MAG_{input})/MAGerr$ as a function of magnitude. As noted by
Benitez \etal~(2004), $MAG_{auto}$ gives better results for
magnitudes, but for photometric redshifts we are interested in good
colors as deep as possible.

\begin{figure}
\plottwo{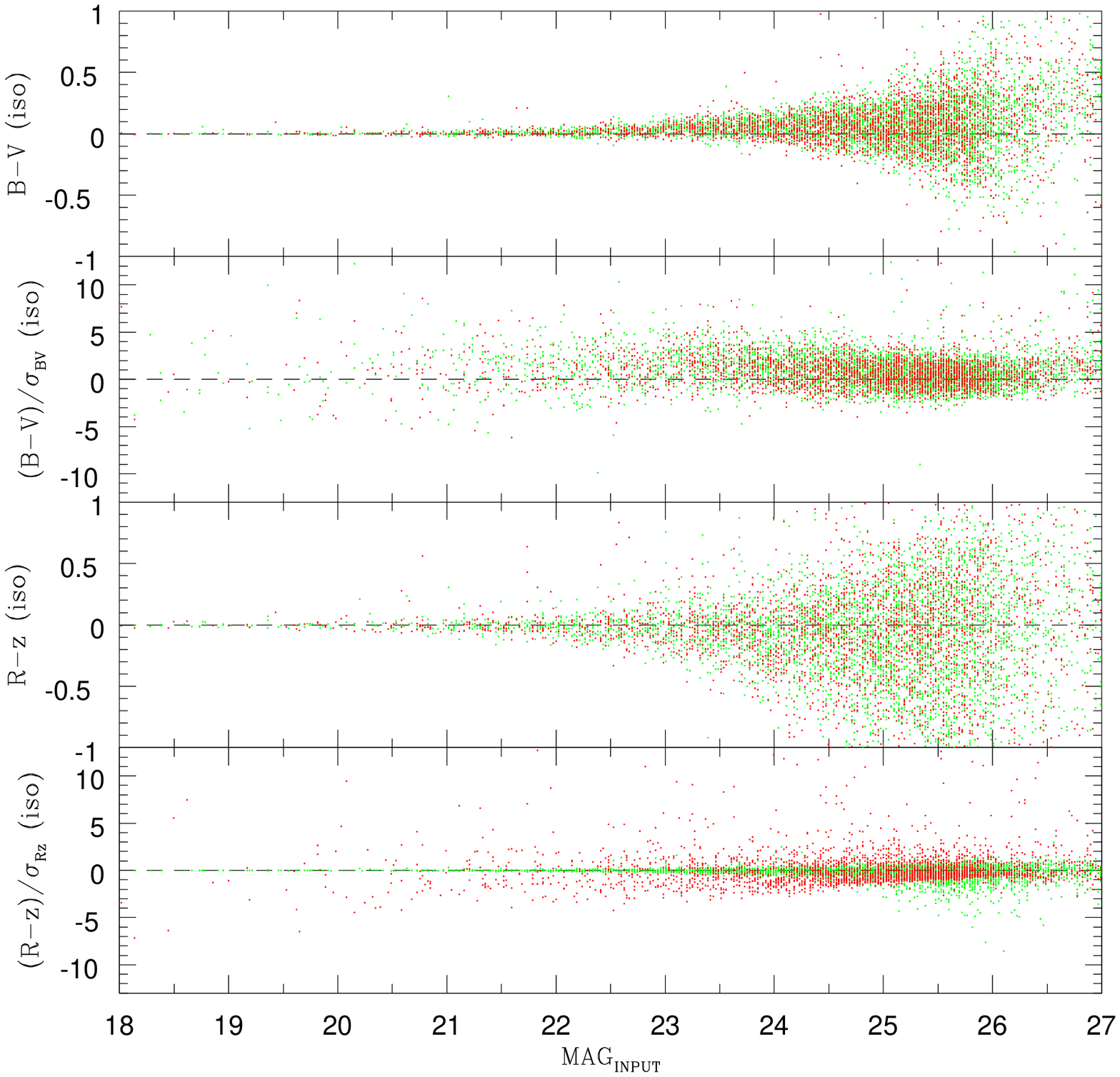}
	{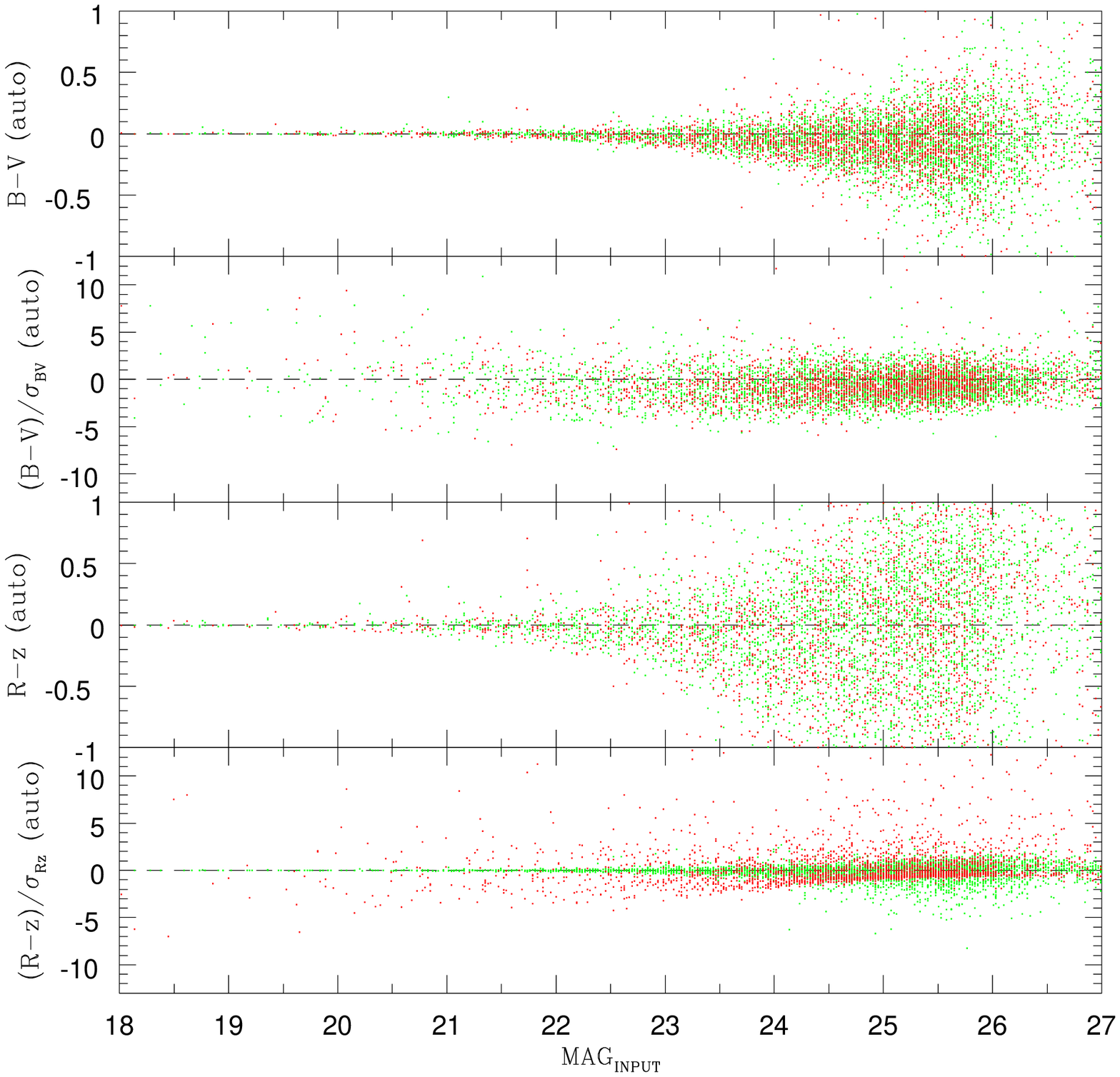}
\caption{Color errors of synthetic De Vaucouleurs (red) and
  exponential disk (green) galaxies added to DLS images.  Left panels
show the SExtractor $MAG_{ISO}$ results while right panels indicate
comparisons to SExtractor $MAG_{AUTO}$.  The four panels in each
column show different color combinations.
\label{fig-colorerr}}
\end{figure}

\begin{figure}
\plotone{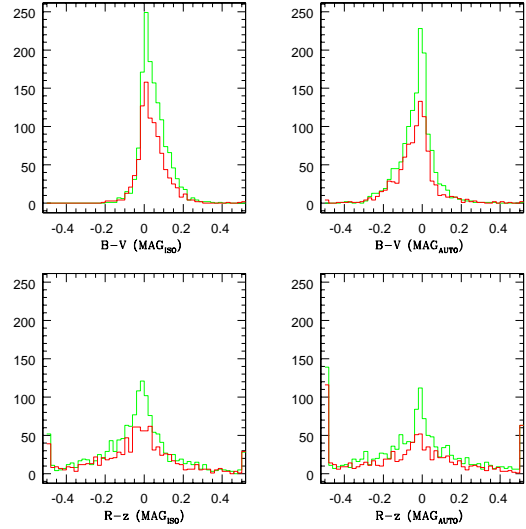}
\caption{Distribution of colors derived from $MAG_{ISO}$ and
  $MAG_{AUTO}$ for zero-color synthetic De Vaucouleurs (red) and
  exponential disk (green) galaxies added to DLS data.  Here we show
  galaxies brighter than $24.5^m$ which corresponds to $\sn\sim10$ in
  BVR, but goes down to $\sn\sim3$ in z.  The edge bins indicate the
  number of objects out of the limits of the plot.
\label{fig-cc}}
\end{figure}

Figures~\ref{fig-colorerr} and \ref{fig-cc} 
show the distribution of {\it color} errors,
which, for photometric redshifts, are more important than magnitude
errors.  Again, $MAG_{iso}$ is on the left and $MAG_{auto}$ on the
right.  The systematic magnitude errors tend to cancel when
considering colors, and $MAG_{iso}$ is now slightly better.  It is
important to note that the errors in magnitude errors are not driven
by faint galaxies, and that in fact the discrepancies between real and
estimated colors errors are significantly worse for bright
objects. 
%Figure~\ref{fig-colors23} shows the cumulative fraction of
%objects with $\Delta_{COLOR}/COLORerr_{iso}$ for galaxies subdivided
%according to magnitude. Galaxies fainter than $23^m$ are represented
%by a short dashed line, while the cumulative fraction for brighter
%galaxies is shown by short-long dashed lines. As before, De
%Vaucouleurs are shown in red, and exponential disk galaxies are in
%green.

\begin{figure}
\plottwo{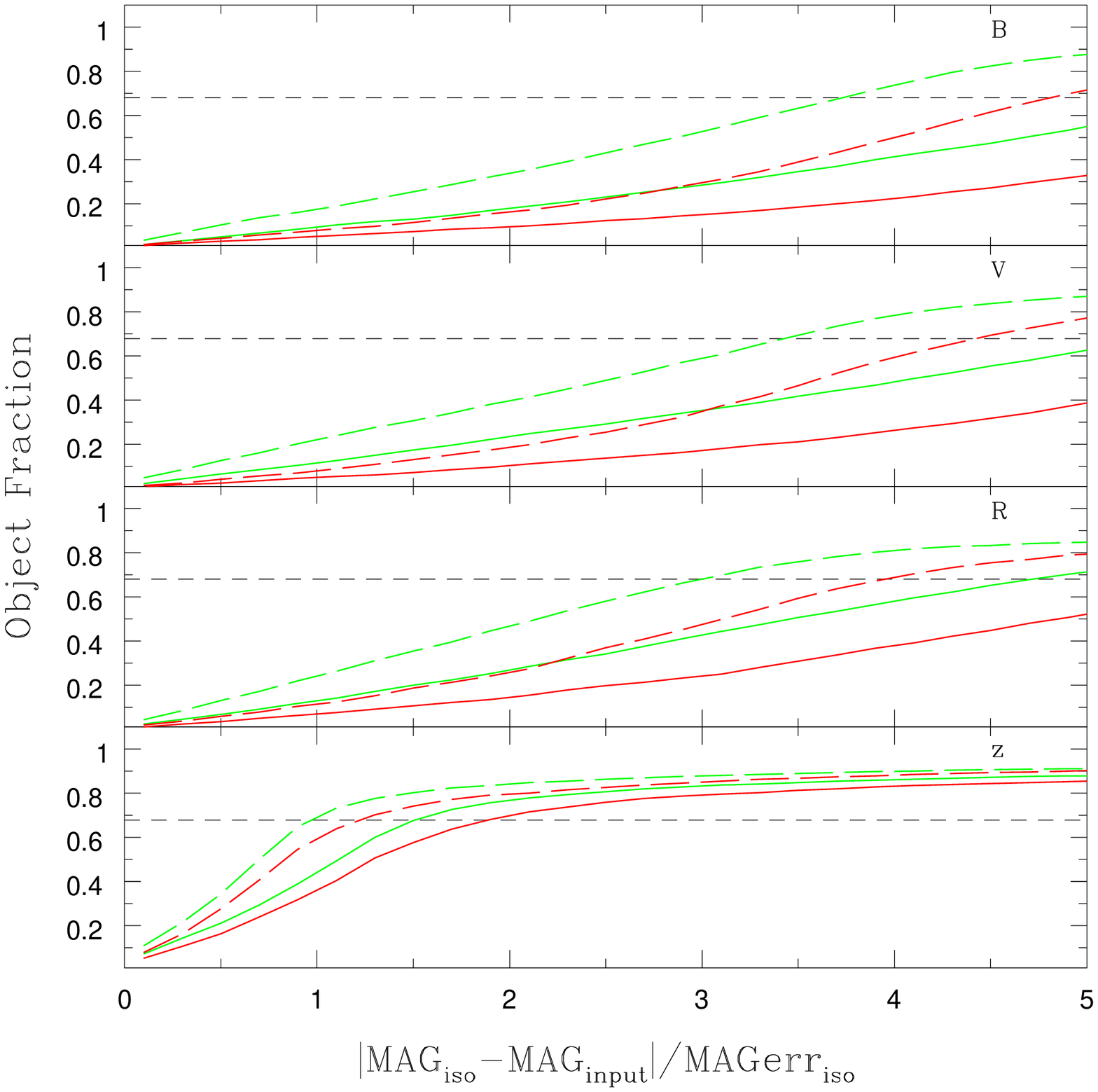}
	{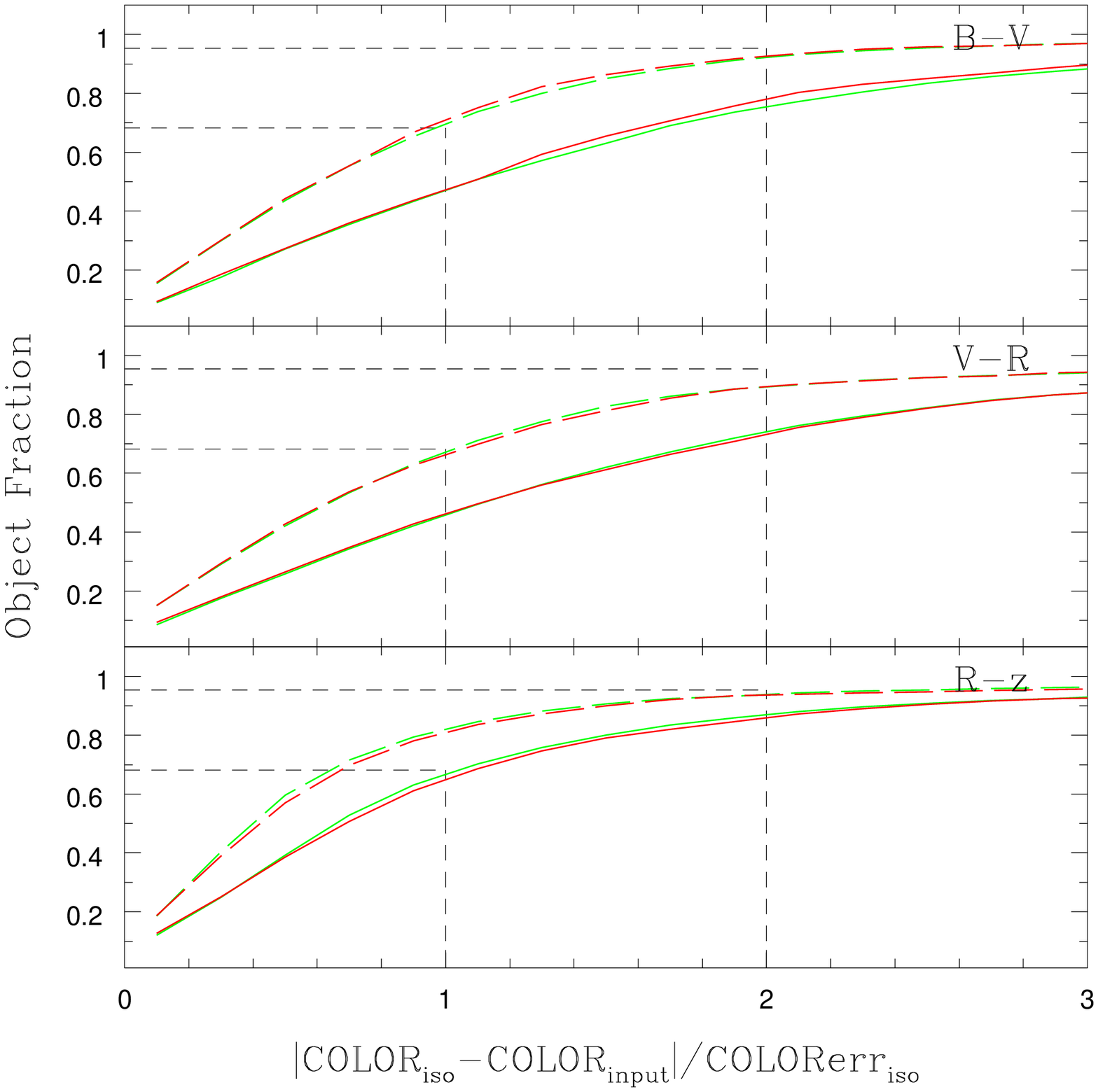}
\caption{Cumulative fraction of objects as function of
  $\Delta_{MAG}/MAGerr_{ISO}$ and $\Delta_{COLOR}/COLORerr_{ISO}$. Red
  lines represent galaxies with a De Vaucouleurs light profile, and
  green lines represent galaxies with an exponential disk. The dashed
  lines indicate the cumulative fraction after an ad hoc increase in
  the measured magnitude errors. The augmented errors guarantee that
  $\sim68(95)\%$ of the galaxies have colors within $1(2)\sigma$.
  A much larger increase would be needed in order to have
  $\sim 68(95)\%$ of galaxies with measured magnitudes within
  $1(2)\sigma$.  
\label{fig-frac}}
\end{figure}

In summary, $MAG_{iso}$ gives slightly more precise colors at a given
magnitude.  This translates to more galaxies being detected above a
given S/N threshold, providing another benefit.  However, for either
$MAG_{auto}$ or $MAG_{iso}$, the error estimates provided by
SExtractor are optimistic, especially at the bright end.  The solid
lines in Figure~\ref{fig-frac} show the cumulative fraction of objects
as a function of magnitude and color error, normalized by the nominal
error from SExtractor.  Much less than $68(95)\%$ of the galaxies have
actual errors within the nominal 1(2)$\sigma$ magnitude error.  Actual
color errors are closer to nominal, but still optimistic.  (Caveat:
unlike most real galaxies, the simulated galaxies had zero color.)
From this analysis we determine an ad hoc correction to the magnitude
errors estimated by SExtractor: we first multiply $MAGerr_{iso}$ by
$1.5$, and then add in quadrature an error of $0.02^m$. The dashed
lines in both panels of Figure~\ref{fig-frac} show the results of this
correction.  This single correction puts the 68th and 95th percentiles
of all the color distributions in the correct place, with the
exception of the 68th percentile of $R-z$ color.  This adjustment to
the magnitude errors should in principle depend on galaxy color, but
we found that variations about this correction made little difference
in the results.

We performed all the real-data tests in this paper with both
$MAG_{iso}$ and $MAG_{auto}$.  The differences in the results were
very minor, except that more galaxies were detected at a given S/N
with $MAG_{iso}$, and about 20\% more survived the ODDS cut with
$MAG_{iso}$.  We therefore adopt $MAG_{iso}$ for the remainder of this
paper.

Another factor to consider is the quality of the survey's photometric
calibration, which was determined by observations of standard stars in
Landolt's (1992) fields during photometric nights.  The R and V DLS
bands are very similar to Landolt's filter transmissions and yield
accurate calibration. The DLS B-band however differs significantly
from Landolt's and requires a color term correction which decreases
the accuracy of calibration in this band. Also, the DLS z-band
photometry derived from Sloan Digital Sky Survey standards (Smith
\etal~2002) is also not as good as R and V. For this reason we add an
extra $0.01^m$ to the magnitude error measurements in B and z
bands. 

%-----------------------------------------------------------------%

\subsection{Template Optimization}\label{tempopt}

\begin{figure}
\plotone{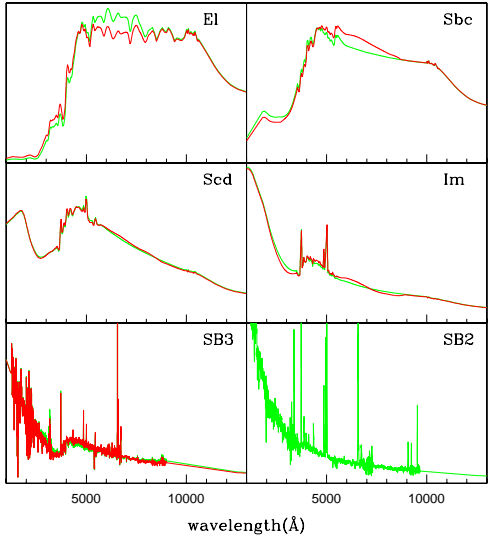}
\caption{Optimized \sed~templates are shown in red, and the original
templates are in green. SB2 template was kept unchanged.
\label{fig-seds}}
\end{figure}

We use spectroscopic redshifts and the DLS photometry to empirically
correct the \bpz~set of templates and to test our filter+instrument
response knowledge with the methodology described in Ilbert \etal
2007. We find optimized templates for El, Sbc, Scd, Im, and SB3 \sed
s. The SB2 template was left unchanged because there were not enough
galaxies of this type to fit a correction. The biggest modifications
were found for the El \sed, which shows a less strong 4000\AA~ break
in the optimized template; and for the Sbc \sed, which has a stronger
4000\AA~ break than in the original BPZ template (See
Figure~\ref{fig-seds}). Because most of our galaxies are at low
redshift, we cannot constrain the longest and shortest SED wavelengths
and therefore we force them to agree with the initial templates.

%-----------------------------------------------------------------%

\subsection{Comparison with Spectroscopic Data: SHeLS Survey}\label{shels}

The SHeLS survey has a limiting magnitude of $R=20.3$, so that the DLS
photometry, which is complete to about five magnitudes fainter, is
very high \sn.  Being a bright magnitude-limited survey, SHeLS
contains overwhelmingly low-redshift ($z<0.6$) galaxies.  However, our
subsample of 1,000 was chosen to provide a nearly uniform redshift
distribution so that characterization accuracy would be roughly
redshift-independent. At a given redshift, selection was random.

\begin{figure}
\plotone{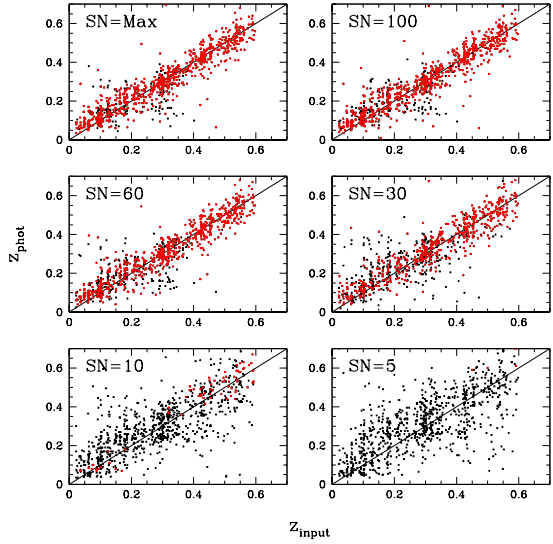}
\caption{$\zp-\zs$ scatter-plot for 860 objects with spectroscopic
 redshifts from the SHeLS survey in the DLS survey. Top-left panel
 shows the results when the maximum \sn~photometry is used.  The five
 other panels (from top-right to bottom=right) show the results with
 progressively greater photometry noise: $\sn=100,60,30,10,5$.  See
 Table~\ref{tab:shelsdata} for statistics.
\label{fig-shelsdatasn}}
\end{figure}

\begin{figure}
\plotone{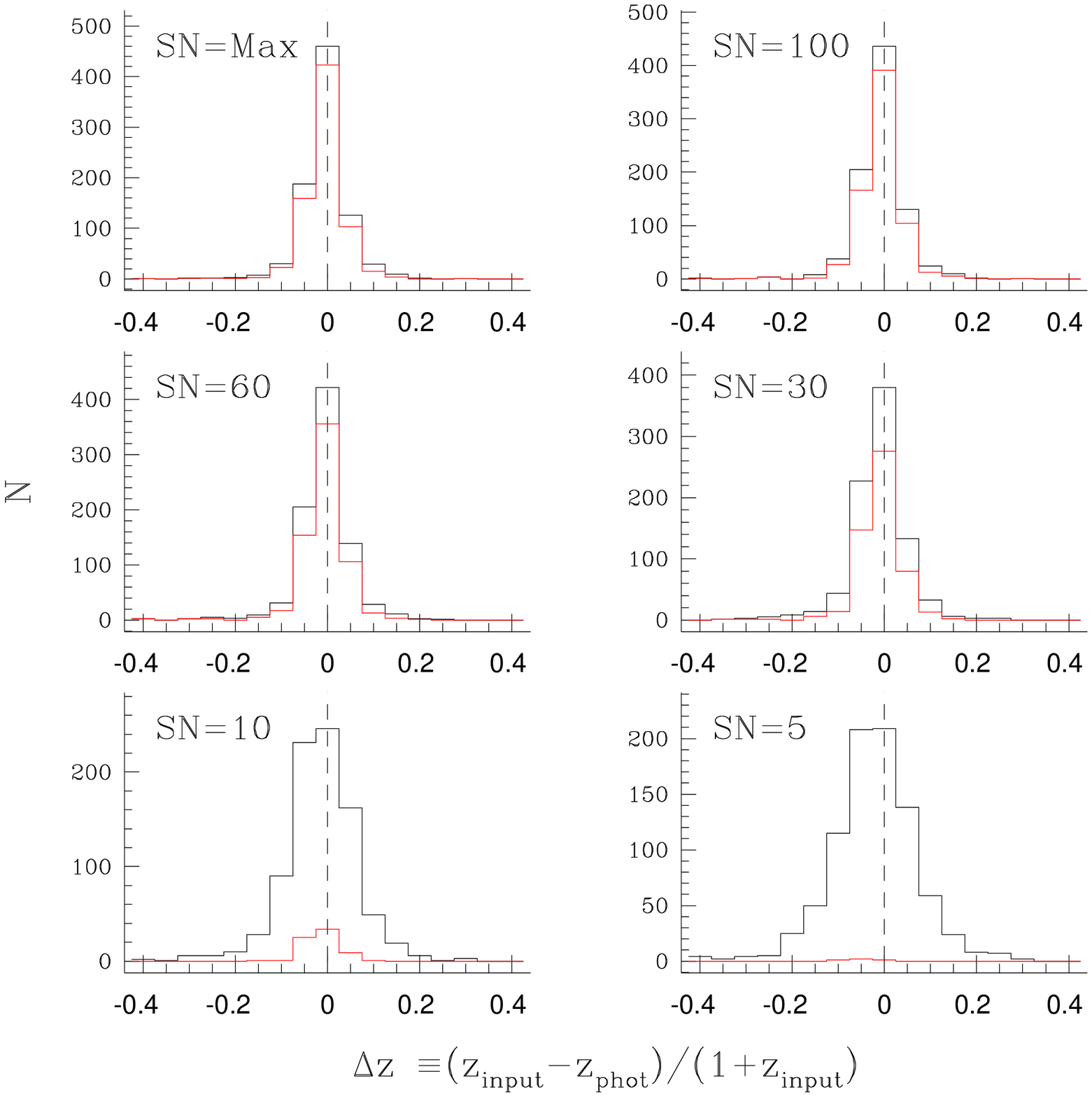}
\caption{Histogram of $\delta z$ for for objects shown in Figure~\ref{fig-shelsdatasn}.
\label{fig-shelsdatasndz}}
\end{figure}

We further cut the sample, requiring $\sn>100$ in the R band, and
excluding objects in exclusion zones around bright stars, or with
saturated pixels in any band, or with SExtractor $flags\ge4$
(compromised photometry).  The final sample contains 860 galaxies.
The top left panels of Figures~\ref{fig-shelsdatasn} and
~\ref{fig-shelsdatasndz} show the $\zp-\zs$ scatter-plot, and $\dz$
distribution for the maximum \sn~ photometry. The distribution of
galaxy types assigned by BPZ to this spectroscopic sample is in
agreement to the type distribution of all galaxies at $R=20\pm0.5^{m}$
in the entire DLS survey, suggesting that the spectroscopic sample is
representative of galaxies at this magnitude.

The SHeLS sample is expected to show evidence of template noise and
have somewhat higher $\sigma_{\dz}$ than the bright end of SIM3, and
this is in fact observed. Objects with $\sn>100$ in SIM3 have
$\sigma_{\dz}=0.037$, and $89.4\%$ of the galaxies have $ODDS>0.9$
with $\sigma_{\dz}=0.026$. For the SHeLS survey, $\sigma_{\dz}=0.050$,
and $85.6\%$ have $ODDS>0.9$ with $\sigma_{\dz}=0.044$. The difference
suggests a template noise of $\sigma_{\dz}\sim0.035(1+z)$ which is
smaller than the $0.065(1+z)$ estimated by Fernandez-Soto \etal~(1999)
for galaxies in the Hubble Deep Field, but expected given the much
lower redshift of galaxies in the SHeLS survey.

\begin{figure}
\plottwo{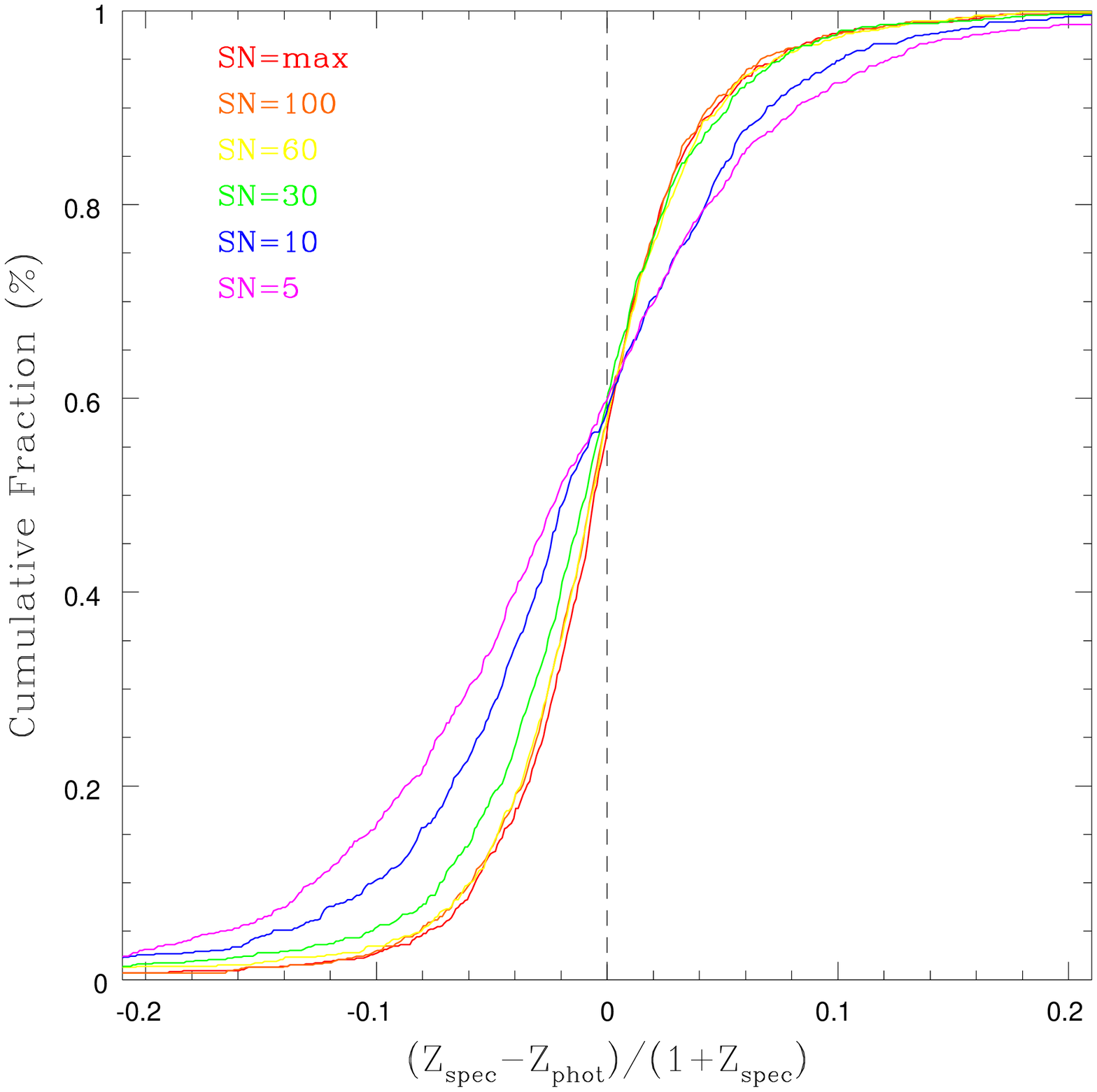}
	{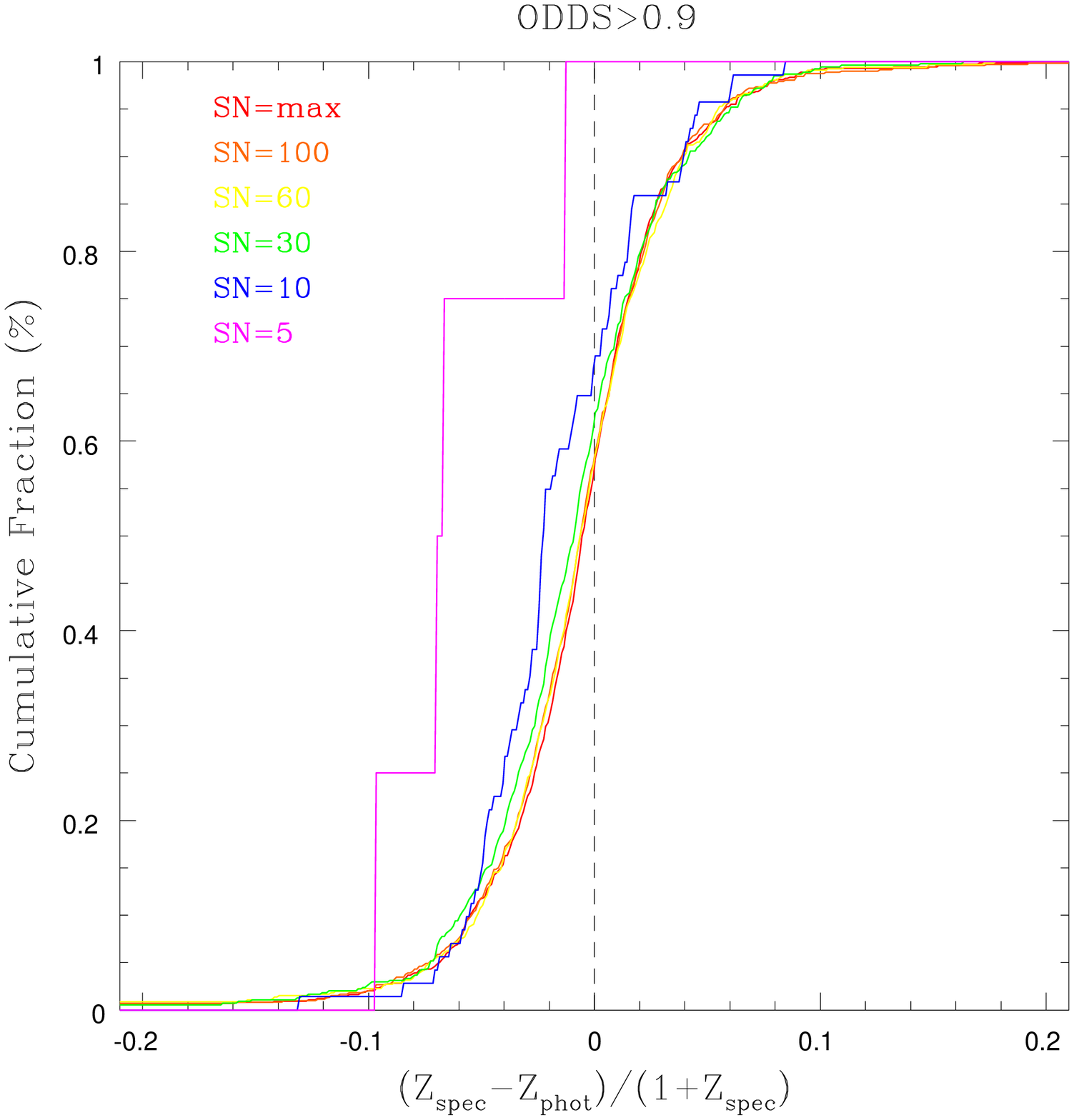}
\caption{Cumulative fraction of objects as a function of $\dz$.  Red
  lines shows the cumulative fraction for maximum \sn~photometry in
  the SHeLS survey; orange shows the fraction when objects are noised
  up simulate $SN=100$ in BVRz; and so on.  Right panel shows all
  galaxies, and left panel shows galaxies with $ODDS>0.9$. %Note that
%  only 2 galaxies have $ODDS>0.9$ in the $\sn=5$ simulation (magenta),
%  and only 75 galaxies have $ODDS>0.9$ in the $\sn=10$ simulation
%  (blue).
\label{fig-shelsfrac}}
\end{figure}

We now degrade the photometry successively to $\sn=100,60,30,10,5$ in
all bands. If a galaxy has, for example $\sn=50$ in the B band, its
magnitude and magnitude error are left unchanged in this band for the
simulations with $\sn=100$ and $\sn=60$, but noise is added to the
other ones. The $\zp-\zs$ scatter-plots are shown in
Figure~\ref{fig-shelsdatasn}. $\dz$ distributions are shown in
Figure~\ref{fig-shelsdatasndz}, and cumulative fraction as a function
of $\dz$ is shown in Figure~\ref{fig-shelsfrac}. Statistics in
different \sn~regimes are presented in Table~\ref{tab:shelsdata}.
The trends with S/N which were observed in the simulations are
reproduced here.  

Because the magnitude prior remains tight despite the \sn~degradation,
we observe lower $\sigma_{\dz}$ at the low \sn~end of the SHeLS
simulations than is observed for SIM2 at the same \sn. At $\sn=10$,
$\sigma_{\dz}=0.080$, and $8.3\%$ of galaxies in the SHeLS survey have
$ODDS>0.9$, while $\sigma_{\dz}=0.121$, and $6.4\%$ of the have
$ODDS>0.9$ for the SIM2 galaxies.

The effectiveness of the ODDS cut is again evident.  The fraction of
galaxies passing this cut at low S/N is less than in SIM3 because the
data here are uniformly at low S/N, whereas for SIM3 the given S/N is
a lower limit.  The fraction with $ODDS>0.9$ at low S/N is more
directly comparable with, and more consistent with, the fractions in
SIM2, which were also at constant S/N.
%Only 2 galaxies have $ODDS>0.9$ in the $\sn=5$ simulation, and
%only 75 galaxies have $ODDS>0.9$ in the $\sn=10$ one.

%-----------------------------------------------------------------%

\subsection{Comparison with Spectroscopic Data: CFGRS Survey}\label{cfgrs}

\begin{figure}
\plottwo{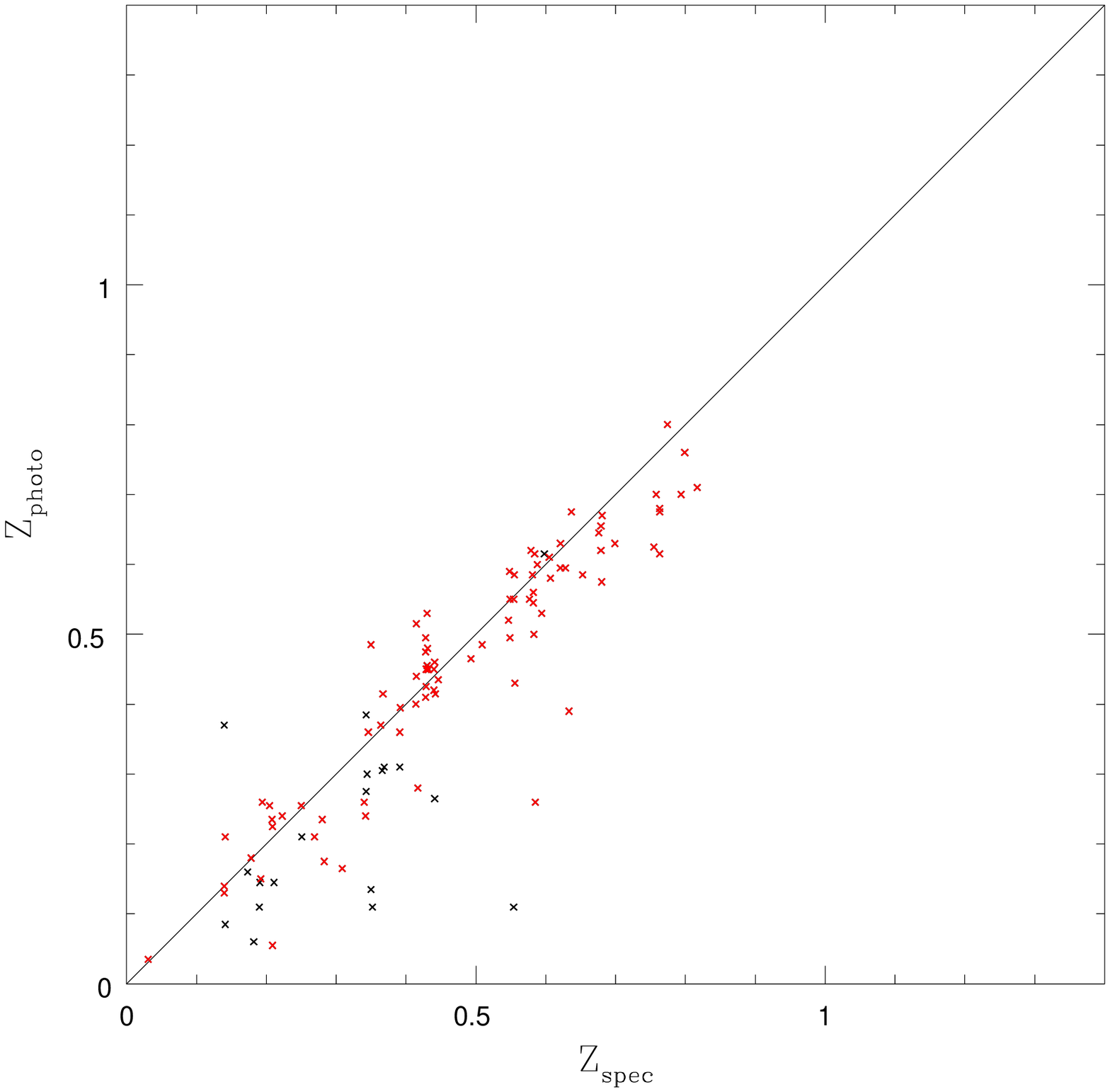}
	{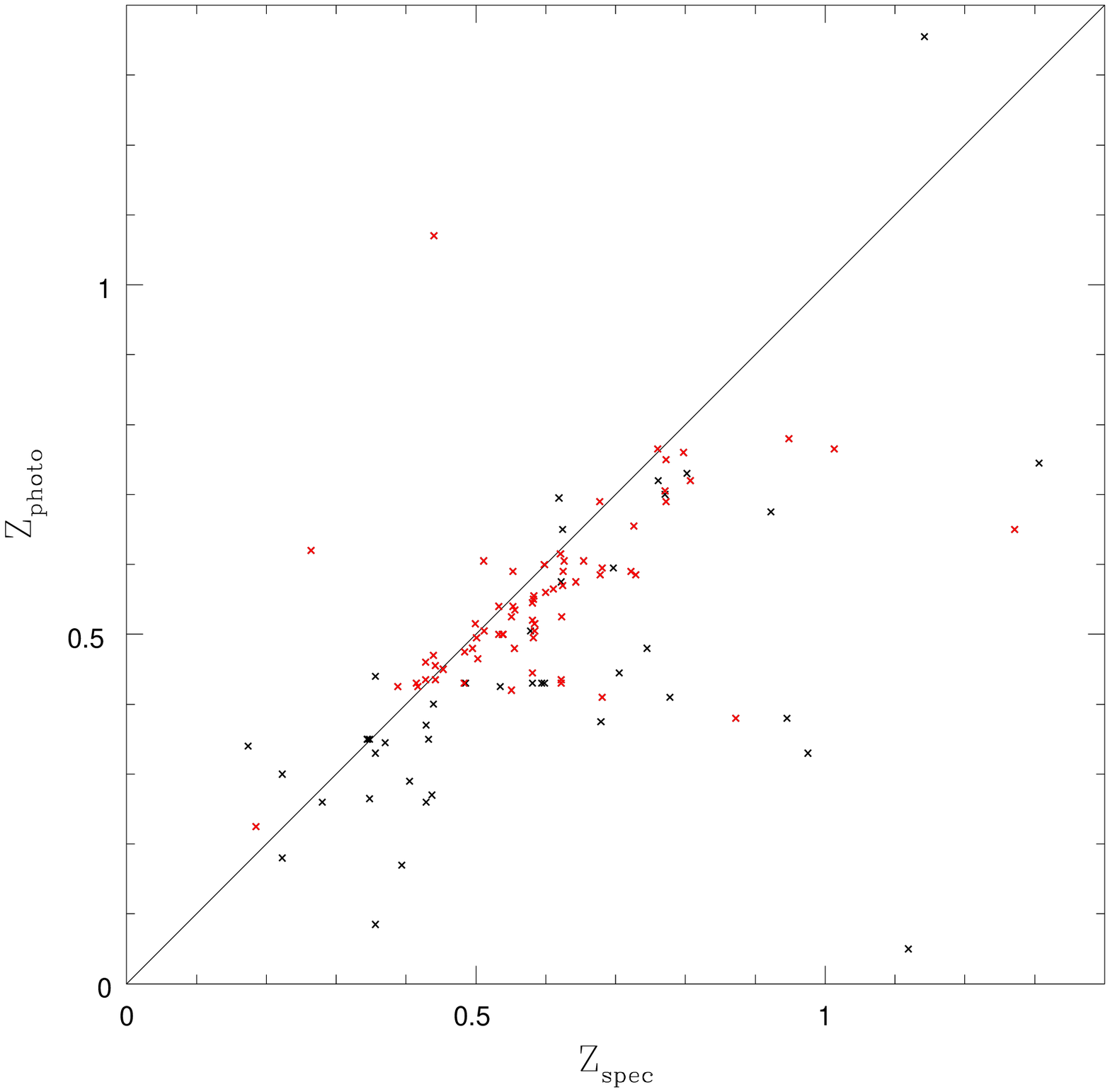}
\caption{$\zp-\zs$ scatter-plot for 222 objects with 
  spectroscopic redshifts from the CFGRS survey. The sample was
  subdivided in two: left, $\sn>106$; right,$\sn<106$. In red, we show
  objects with $ODDS>0.9$.
\label{fig-cfgrs}}
\end{figure}

The CFGRS (Cohen \etal~1999) survey is about $2^m$ deeper than SHeLS
and therefore the DLS photometry is not as high \sn.  We select
galaxies with quality=1 (multiple spectral features, Cohen \etal~1999)
spectroscopic redshifts and divide the data in 2 equally sized
subsamples of 111 galaxies each: one with galaxies of photometric
$\sn(R)>106$, and another with $\sn(R)<106$.  Note that the
signal-to-noise in the low
\sn~ sample is still fairly high, with 28 being the lowest value, and
a median of 69, but the difference in the quality of photometric
redshifts is clear.  Figure~\ref{fig-cfgrs} shows the $\zp-\zs$
scatter-plot for the two sub-samples.  For the high \sn~sample,
$\dz=0.027\pm0.084$, and $\dz=0.021\pm0.060$ if we exclude 1
catastrophic outlier with $|\dz|>0.5$.  For the lower
\sn~sample, $\dz=0.033\pm0.166$, and $\dz=0.041\pm0.095$ if we exclude
2 objects with $|\dz|>0.5$. However this includes the effect of
different redshift ranges. To isolate the
\sn~effect, we compute results using only galaxies between
$0.4<z<0.9$, where both samples have a significant density of sources.
For the high \sn~sample we find $\dz=0.020\pm0.056$, and for the lower
\sn~sample, we find $\dz=0.034\pm0.073$. No objects with $|\dz|>0.5$
are found in this redshift range.

%-----------------------------------------------------------------%
%-----------------------------------------------------------------%

\section{Selection in Galaxy Type and Redshift Range}

\begin{figure}
\plotone{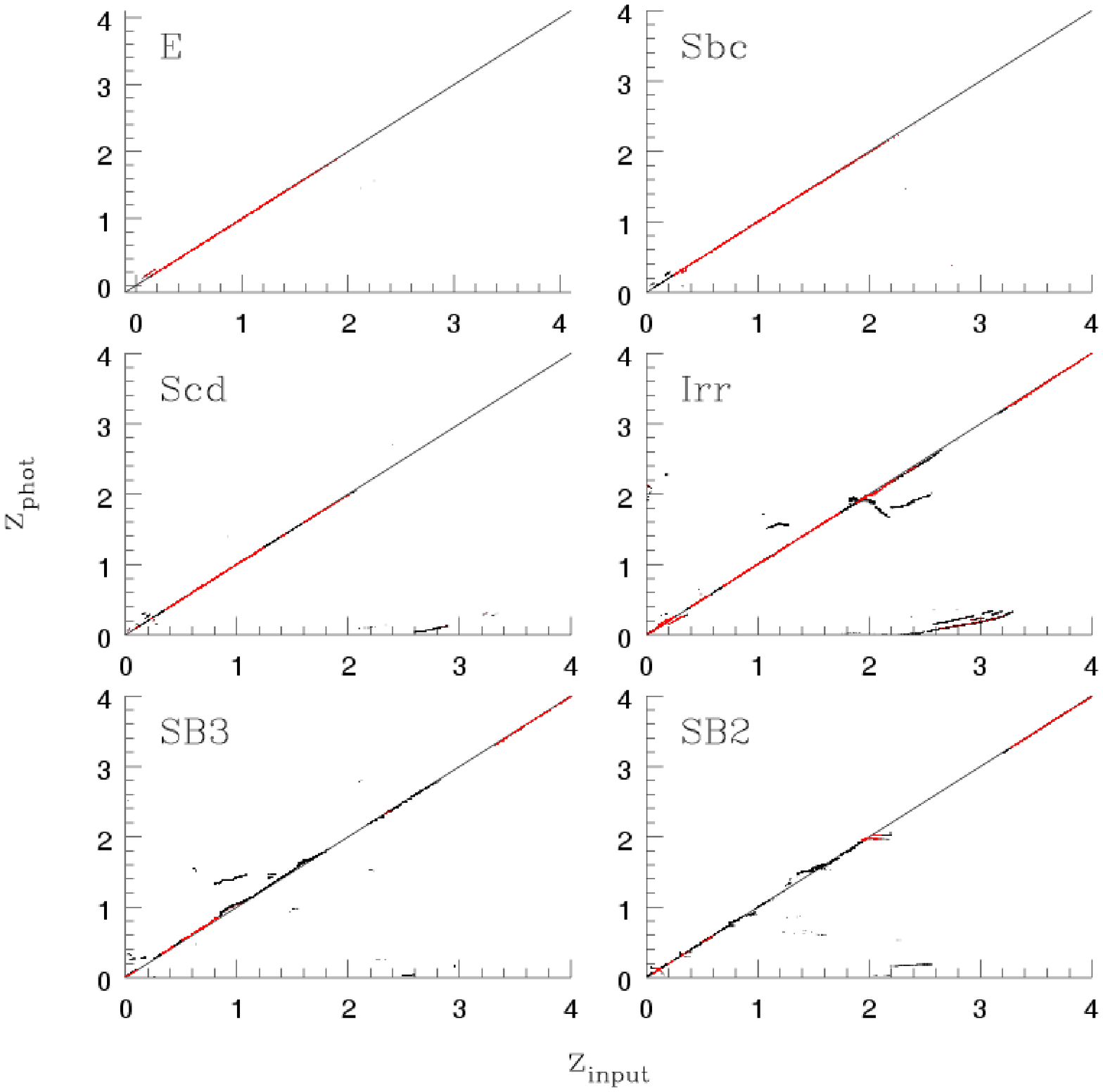}
\caption{$\zp-\zs$ scatter-plot for SIM1 subdivided according to
  \bpz~galaxy type ($T_B$). Galaxies with $ODDS>0.9$ are in red.
\label{fig-sim1typ}}
\end{figure}

\begin{figure}
\plotone{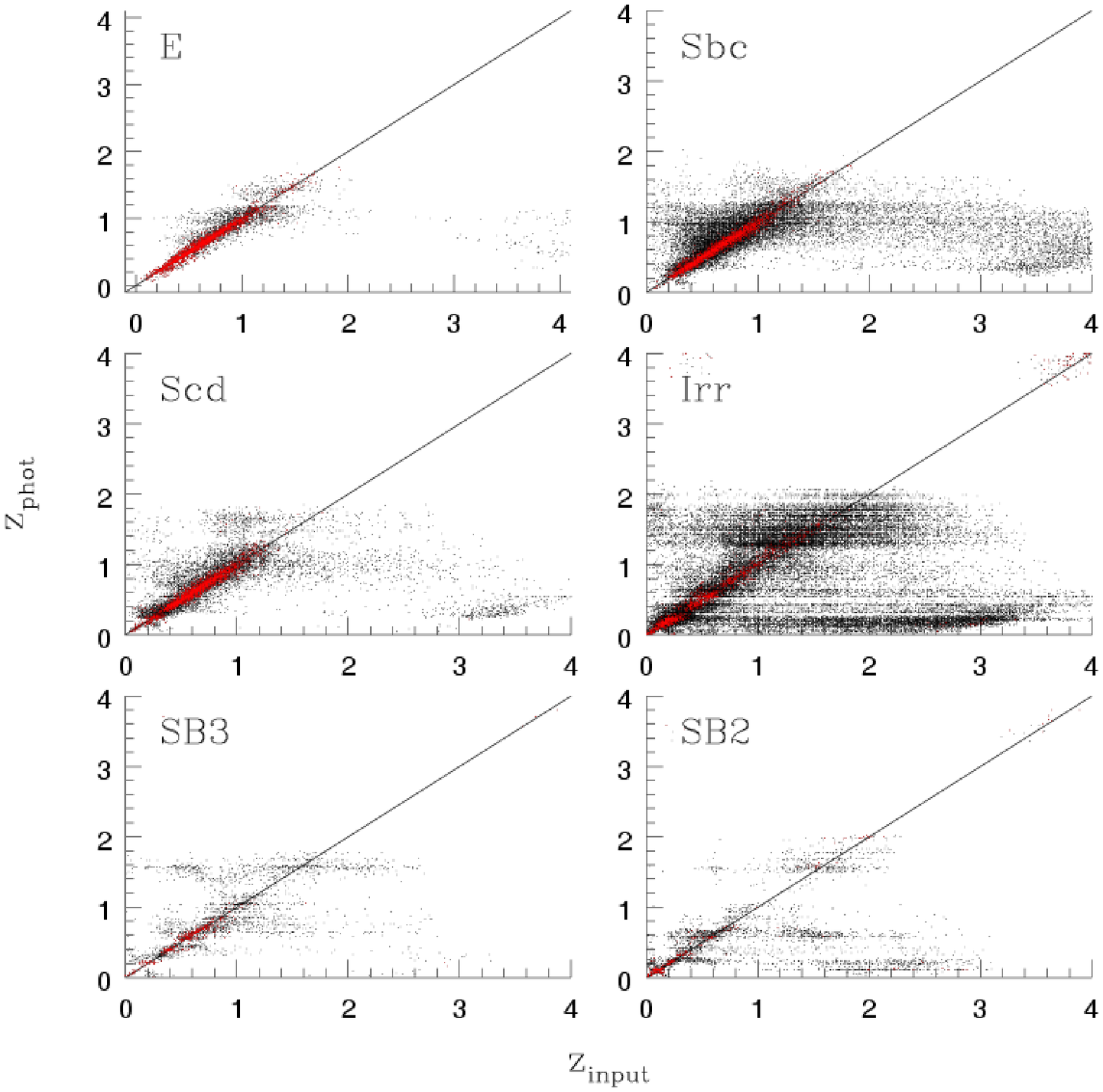}
\caption{$\zp-\zs$ scatter-plot for SIM3 subdivided according to
  \bpz~galaxy type ($T_B$). Galaxies with $ODDS>0.9$ are in red. 
\label{fig-sim3typ}}
\end{figure}

\begin{figure}
\plotone{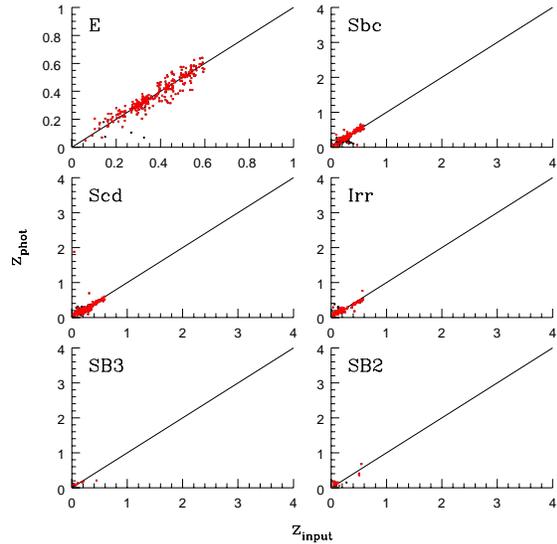}
\caption{$\zp-\zs$ scatter-plot subdivided according to \bpz~galaxy
  type ($T_B$) for 860 objects with spectroscopic redshifts from
  SHeLS. Galaxies with $ODDS>0.9$ are in red.
\label{fig-datatyp}}
\end{figure}

Figure~\ref{fig-priors2} suggests that faint Irr/SB2/SB3 galaxies are
often misclassified as Sbc/Scd. In this section we explore dependence
on type in more detail.  Figures~\ref{fig-sim1typ},
~\ref{fig-sim3typ}, and ~\ref{fig-datatyp} show the $\zp-\zs$
scatter-plot as a function of inferred \bpz~galaxy type ($T_B$) for
SIM1, SIM3, and the SHeLS galaxies respectively.  Ellipticals form the
tightest relation, while the redshift of irregular galaxies show a
scatter more than twice as large.  Figure~\ref{fig-sim3typ} shows that
some of the scatter in ellipticals must be due to misclassifications,
because there are no E-type galaxies at $z\sim3-4$ in the simulations.

\begin{figure}
\plotone{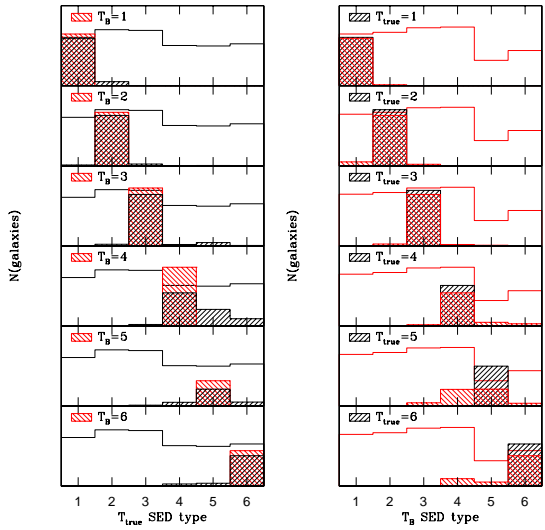}
\caption{Input spectral type ($T_{true}$) {\it vs.} output \bpz ~type
  ($T_B$) for SIM3 galaxies with $S/N>30$. Types are: $1-E$; $2-Sbc$;
  $3-Scd$; $4-Irr$; $5-SB3$; $6-SB2$.  In the left panels we select
  galaxies by $T_B$ (shaded red) and then look at their $T_{true}$
  distribution (shaded black). The unshaded black histogram is the
  same in all plots and indicates the $T_{true}$ distribution. In the
  right panels, we select galaxies by their true type (shaded black)
  and then look at the $T_B$ distribution (shaded red). The unshaded
  red histogram indicates the $T_{B}$ distribution.
\label{fig-types}}
\end{figure}

\begin{figure}
\plotone{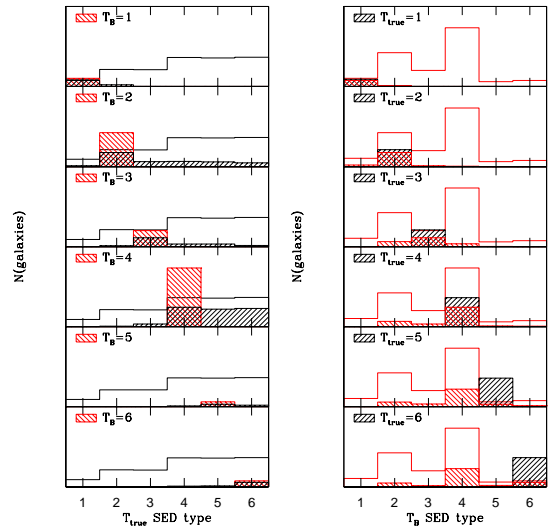}
\caption{Same as Figure~\ref{fig-types} but including galaxies with
  very low \sn~(all detections). Black represents true galaxy types
  (input) and red indicates the $T_B$ classification.
\label{fig-typesall}}
\end{figure}

We look at type misclassification in SIM3 directly in
Figures~\ref{fig-types} and ~\ref{fig-typesall}.  The left column of
panels shows the $T_B$ distribution for each of the true input types,
with the true type distribution overlaid like a diagonal matrix in red
to guide the eye.  The right column of panels shows the true type
distribution for each of the inferred types, with the inferred type
distribution overlaid in red to guide the eye.  The overall
distribution of inferred (true) types is shown by the unshaded
histogram which is repeated in each panel in the left (right) column.
Figures~\ref{fig-types} shows galaxies with $\sn\ge30$ or $R\le23$.
For example, the fourth panel down in the left column shows that
galaxies classified as $T_B=4$ (Irr), have in fact almost the same
probability of being of types 4, 5 or 6 (irregular or starburst).
Likewise, starburst galaxies tend to be misclassified at irregulars
even at high \sn.

The types in decreasing order of reliability are E, Sbc, Scd, Irr,
SB3, and SB2.  Type reliability translates to redshift reliability,
because type misclassification usually implies a large, if not
catastrophic, redshift error.  These figures also demonstrate that
although the ODDS cut appears to lose many high high-redshift galaxies
and shrink the usable redshift range, in fact most of the
``high-redshift'' galaxies lost were type misclassifications, and
therefore unreliable redshifts.  Although the loss of these
``high-redshift'' galaxies is painful if one wants as large a redshift
range as possible, it is necessary if one wants the sample to be
reliable.

In Figure~\ref{fig-typesall} we extend the analysis to lower
\sn~galaxies, and include all ``detected'' galaxies. The
rate of misclassification is much higher. The insertion of these
objects in the sample creates new types of misclassification. For
example, a fraction of type 1 (E) galaxies is assigned $T_B=2$ and
vice-versa. Also, a significant fraction of types 4, 5, and 6
(irregular and starburst) are classified as types 2 or 3 (spirals).

%-----------------------------------------------------------------%
%-----------------------------------------------------------------%

\section{Summary and Discussion}

We have examined the dependence of photometric redshift performance on
photometric S/N, using both simulations and data.  For concreteness,
we have used the DLS filter set, but the general trends should apply
to any filter set.  As a reminder, SIM2 simulated galaxies at a range
of magnitudes drawn from the DLS photometry, but at a series of
constant S/N levels, while SIM3 strongly couples magnitude and S/N as
they are in the DLS photometry.  Thus, {\it bright} is distinct from
{\it high S/N} in SIM2 and in the noise-augmented SHeLS data because
{\it bright} implies a more effective magnitude prior.  An additional
distinction between SIM3 and the other cases is that in SIM3 a given
S/N cut is performed in R, and for most galaxies that implies a lower
S/N in the other bands. For SIM2 and noise-augmented SHeLS data, a
given S/N describes each galaxy in each filter.

We therefore expect the smallest $\sigma_{\dz}$ for very high S/N in
SIM3, where the high S/N galaxies automatically have a tight magnitude
prior.  This is what is observed, $\sigma_z=0.031$ (0.037) for
$S/N>250$ (100) in SIM3.  Degeneracies in color space determine this
performance limit, which is therefore highly filter-set dependent.
However, it sets a baseline for what follows.  At $S/N=100$ in the
SHeLS data, $\sigma_{\dz}$ is about 35\% larger than this baseline,
suggesting a cosmic variance or template noise component of
$\sigma_{\dz}=0.035(1+z)$.  For SIM2, $\sigma_{\dz}$ is also about
32\% larger than this baseline, presumably due to the looser magnitude
priors on average.  The deeper the survey, the less effective the
magnitude prior, but performance is still quite good at this high S/N.

From this baseline, lowering the S/N smoothly increases $\sigma_{\dz}$
in SIM3, by 30--50\% at each S/N step in Table~\ref{tab:sim3} until
$\sigma_{\dz}$ is no longer trustworthy due to the clipping at
$|\dz|>0.5$.  SIM2 degrades a bit more slowly due to its higher
baseline $\sigma_{\dz}$.  The noise-augmented SHeLS data degrades even
more slowly, because magnitude priors always remain tight.  Although
$\sigma_{\dz}$ looks reasonably good even at $S/N=5$ for the degraded
SHeLS data, we expect SIM3 to be more representative of true
performance for this reason.

%somewhat tighter at the lowest S/N because, unlike SIM3, each galaxy
%has that S/N in {\it each} filter.  This is enough to overcome the
%opposing effect that no galaxy in SIM2 has {\it more} than the labeled
%S/N. 

SIM3 indicates that without an ODDS cut, $S/N=17$ in R is likely to be
the lowest acceptable S/N for reasonable photometric redshift
performance ($\sigma_{\dz}=0.1$) in a survey with the DLS
specifications (filter set and depth).  A shallower survey may be able
to go to lower S/N because the magnitude prior remains helpful to
lower S/N in such a survey.  In fact, the bright spectroscopic sample
has $\sigma_{\dz}<0.1$ even at $S/N=5$, although we caution that this
means $S/N=5$ in {\it each} filter.  If we impose an ODDS cut rather
than an S/N cut, $ODDS>0.40$ cut yields twice as many galaxies for the
same $\sigma_{\dz}$ as the $S/N>17$ cut in R.  Alternatively, survey
users could use ODDS to decrease $\sigma_{\dz}$ while sacrificing
galaxy counts; an $ODDS>0.9$ cut yields $\sigma_{\dz}=0.04$ averaged
over all S/N.

We caution that there are some unmodeled effects which, if included,
would result in a larger $\sigma_{\dz}$.  First, template noise is not
included in the simulations.  $\sigma_{\dz}$ is larger in the SHeLS
data than in SIM3 for $S/N>60$, which we attribute to template noise.
Template noise becomes less important at lower photometric S/N, but
the template noise in the SHeLS data may be artificially low.  The
templates were originally derived from bright galaxies like those in
SHeLS, and further optimized on the SHeLS sample itself.  A
photometric sample which pushes to higher redshift may thus incur more
template noise, and in fact Fernandez-Soto \etal~(1999) estimates
$\sigma_{\dz}=0.065(1+z)$ for galaxies in the Hubble Deep Field.

Second, because galaxy counts are rising beyond the limiting magnitude
for detection, an additional source of photometry noise must be taken
into account.  A source detected at S/N of a few is much more likely
to be an ``up-scattered'' fainter galaxy than a ``down-scattered''
brighter galaxy.  As pointed out by Hogg \& Turner (1998, hereafter
HT98), this is distinct from Malmquist bias, which is the
over-representation of high-{\it luminosity} galaxies in a flux-limited
sample.  Although the resulting bias can be computed and corrected for
if the galaxy count slope is known, the additional photometric
uncertainty is unavoidable.  In fact, HT98 conclude that ``sources
identified at signal-to-noise ratios of four or less are practically
useless.''  This source of noise was not reproduced in our
simulations, so extrapolation to $S/N<5$ would be extremely dangerous.
Our results for $S/N=5$ are still valid if five is interpreted as the
effective S/N in the presence of this additional source of noise.  For
the no-evolution, Euclidean slope of $q=1.5$, the HT98 formulae
indicate that this requires a detection at $S/N=5.64$.  For $S/N=10$
and higher, the corrections are very small.

In addition to these dependences on S/N, several other lessons can be
drawn:
\begin{itemize}

\item When forecasting photometric redshift performance for a survey,
it is important to include realistic photometry errors.  

\item Estimating photometric redshift performance with spectroscopic
samples can lead to optimistic results if the spectroscopic sample is
not representative of the photometric sample.  If the spectroscopic
sample is brighter, matching the S/N is easily accomplished by adding
photometry noise, but accounting for the larger redshift range of the
photometric sample requires detailed modeling which must account for
cosmic variance.

\item The BPZ $ODDS$ parameter is very effective at identifying
photometric redshifts which are likely to be poor.  An $ODDS$ cut is
more efficient than an S/N cut, because $ODDS$ takes account of the
looser photometry requirements in distinctive regions of color space.
Still, our simulations and artificially noisy data show that of the
galaxies with $ODDS<0.9$, the ones with poor photometric redshifts may
be in the minority.  The tradeoff between $ODDS$ cut and usable
numbers of galaxies must be assessed in light of the specific science
goal.  For example, if the science analysis weights each galaxy by its
photometric S/N, a strict $ODDS$ cut may cut most of the galaxies but
not most of the total weight.  For weak lensing, shape noise limits
the maximum weight of a galaxy, so a strict $ODDS$ cut may cut most of
the weight.  Finally, biases must be considered, as ellipticals are
overrepresented in the set of galaxies with high $ODDS$.  This may not
affect weak lensing but will be important for studies of galaxy
evolution and baryon acoustic oscillations.

\end{itemize}

We also explored cutting in type (as identified by BPZ) and redshift
range.  As expected, ellipticals do better than any other type, but we
found that the $ODDS$ cut was still useful for ellipticals.  As long
as the $ODDS$ cut was being used, other types could safely be used as
well.  Therefore, we recommend cutting on ODDS rather than type.

\acknowledgments

We thank NOAO for supporting survey programs and the CFGRS project for
making data publicly available. DLS observations were obtained at
Cerro Tololo Inter-American Observatory (CTIO) and Kitt Peak National
Observatory (KPNO). CTIO and KPNO are part of the National Optical
Astronomy Observatory (NOAO), which is operated by the Association of
Universities for Research in Astronomy, Inc., under cooperative
agreement with the National Science Foundation.  We also would like to
thank Margaret Geller and Michael Kurtz for providing us with 1,000
SHeLS redshifts, which were observed with Hectospec at the MMT
Telescope.

We  thank Ian Dell'Antonio  and Tony  Tyson for  comments that  led to
improvements to the paper.

%%%%%%%%%%%%%%%
%%% FIGURES %%%
%%%%%%%%%%%%%%%

%-SIM----------------------------------------------------------------%

%\clearpage

%\clearpage

%\clearpage

%-SIM1----------------------------------------------------------------%

%\clearpage

%\clearpage

%-SIM2----------------------------------------------------------------%

%\clearpage

%\clearpage

\clearpage

%-SIM3----------------------------------------------------------------%

%\clearpage

%\clearpage

%\clearpage

%\clearpage

%\clearpage

%\begin{figure}
%\plotone{f14.ps}
%\caption{$N(\zi)$ for successive cuts in \sn. Black: No SN cut; Blue:
%  $SN>10$; Green: $SN>30$; Yellow: $SN>60$; Orange: $SN>100$; Red:
%  $SN>250$. Note that the galaxy mixture also changes according to the
%  priors shown in Figure~\ref{fig-priors}.
%\label{fig-nzinputsn}}
%\end{figure}
%\clearpage

\clearpage

%-Data----------------------------------------------------------------%
%-Measuring Colors----------------------------------------------------%

%\clearpage

%\clearpage

\clearpage

%-Template Optimization-------------------------------------------%

%\clearpage

%-Shels-----------------------------------------------------------%

%\clearpage

%\clearpage

\clearpage

%-CFGRS-----------------------------------------------------------%
 
%\clearpage

%-Galaxy Type-----------------------------------------------------%

%\clearpage

%\clearpage

%\clearpage

%\clearpage

\clearpage

%%%%%%%%%%%%%%%
%%% TABLES  %%%
%%%%%%%%%%%%%%%

%-SIM1 & SIM2-----------------------------------------------------------%
%/home/vem/mypapers/ZPHOT2/table.sim12.tex

\begin{deluxetable}{cccccccc}
\tablewidth{0pt}
\tabletypesize{\footnotesize}
\tablecaption{Galaxies with Fixed Signal-to-Noise (SIM1 \& SIM2)
\label{tab:sim12}}
\tablehead{\colhead{SN} &\multicolumn{3}{c}{$|\dz|\le0.5$} &\multicolumn{4}{c}{$ODDS>0.9$}\\
\colhead{} &\colhead{$|\dz|\le0.5 \over All$(\%)} &\colhead{$\bar{\dz}$} &\colhead{$\sigma(\dz)$} &\colhead{$ODDS>0.9\over All$(\%)} &\colhead{$|\dz|\le0.5~\&~ODDS>0.9 \over ODDS>0.9$(\%)} &\colhead{$\bar{\dz}$} &\colhead{$\sigma(\dz)$}}
\startdata
  Inf (SIM1)  &   95.3 & -0.008 &  0.051 &   53.4 &   99.7 & -0.000 &  0.006 \\
  250 (SIM2)  &   96.0 & -0.005 &  0.042 &   64.2 &   99.8 & -0.000 &  0.009 \\
  100 (SIM2)  &   95.5 & -0.007 &  0.049 &   60.7 &   99.7 & -0.000 &  0.012 \\
   60 (SIM2)  &   94.7 & -0.010 &  0.062 &   54.4 &   99.7 & -0.001 &  0.015 \\
   30 (SIM2)  &   92.5 & -0.014 &  0.085 &   40.6 &   99.9 & -0.002 &  0.020 \\
   10 (SIM2)  &   87.6 & -0.007 &  0.121 &    6.4 &  100.0 & -0.001 &  0.023 \\
    5 (SIM2)  &   84.7 &  0.019 &  0.151 &    1.2 &   99.9 &  0.001 &  0.012 \\
\enddata
%\tablecomments{}
\end{deluxetable}

%-SIM3------------------------------------------------------------------%
%/home/vem/mypapers/ZPHOT2/table.sim3.tex

\begin{deluxetable}{cccccccc}
\tablewidth{0pt}
\tabletypesize{\footnotesize}
\tablecaption{Galaxies in DLS like simulations (SIM3)
\label{tab:sim3}}
\tablehead{\colhead{SN(R)} &\multicolumn{3}{c}{$|\dz|\le0.5$} &\multicolumn{4}{c}{$ODDS>0.9$}\\
\colhead{} &\colhead{$|\dz|\le0.5 \over All$(\%)} &\colhead{$\bar{\dz}$} &\colhead{$\sigma(\dz)$} &\colhead{$ODDS>0.9\over All$(\%)} &\colhead{$|\dz|\le0.5~\&~ODDS>0.9 \over ODDS>0.9$(\%)} &\colhead{$\bar{\dz}$} &\colhead{$\sigma(\dz)$}}
\startdata
$>250$   &  100.0 & -0.001 &  0.031 &   90.9 &  100.0 & -0.001 &  0.021 \\
$>100$   &  100.0 &  0.001 &  0.037 &   89.4 &  100.0 &  0.000 &  0.026 \\
$>60$    &   99.7 & -0.000 &  0.050 &   82.6 &  100.0 &  0.000 &  0.030 \\
$>30$    &   97.8 & -0.004 &  0.076 &   67.0 &   99.6 & -0.001 &  0.036 \\
$>10$    &   89.3 & -0.004 &  0.125 &   23.5 &   99.3 & -0.001 &  0.040 \\
$>5$     &   85.0 &  0.008 &  0.154 &   14.1 &   99.3 & -0.001 &  0.040 \\
All      &   83.6 &  0.018 &  0.170 &   11.9 &   99.3 & -0.001 &  0.040 \\
\enddata
%\tablecomments{}
\end{deluxetable}

%-SHELS-----------------------------------------------------------------%
%/home/vem/mypapers/ZPHOT2/table.shelsdata.tex

\begin{deluxetable}{cccccccc}
\tablewidth{0pt}
\tabletypesize{\footnotesize}
\tablecaption{Galaxies with Spectroscopic Redshifts from the SHeLS Survey
\label{tab:shelsdata}}
\tablehead{\colhead{SN} &\multicolumn{3}{c}{$|\dz|\le0.5$} &\multicolumn{4}{c}{$ODDS>0.9$}\\
\colhead{} &\colhead{$|\dz|\le0.5 \over All$(\%)} &\colhead{$\bar{\dz}$} &\colhead{$\sigma(\dz)$} &\colhead{$ODDS>0.9\over All$(\%)} &\colhead{$|\dz|\le0.5~\&~ODDS>0.9 \over ODDS>0.9$(\%)} &\colhead{$\bar{\dz}$} &\colhead{$\sigma(\dz)$}}
\startdata
Max &   99.9 & -0.005 &  0.050 &   85.6 &   99.9 & -0.006 &  0.044 \\
100 &   99.8 & -0.006 &  0.050 &   83.0 &   99.9 & -0.007 &  0.045 \\
 60 &   99.8 & -0.006 &  0.054 &   76.9 &   99.7 & -0.006 &  0.045 \\
% 50 &   99.7 & -0.008 &  0.055 &   75.1 &   99.8 & -0.007 &  0.044 \\
 30 &  100.0 & -0.012 &  0.061 &   62.9 &  100.0 & -0.010 &  0.046 \\
% 20 &  100.0 & -0.013 &  0.067 &   46.3 &  100.0 & -0.010 &  0.048 \\
 10 &  100.0 & -0.016 &  0.080 &    8.3 &  100.0 & -0.015 &  0.038 \\
  5 &   99.8 & -0.021 &  0.090 &    0.5 &  100.0 & -0.061 &  0.035 \\
\enddata
%\tablecomments{}
\end{deluxetable}

\end{document}